\documentclass[aps, superscriptaddress, showpacs, prl, twocolumn,  floatfix]{revtex4-2}
\usepackage{blindtext}
\usepackage[english]{babel}
\usepackage[dvipsnames]{xcolor}
\usepackage{array}
\usepackage{subfigure}
\usepackage{float}
\usepackage{mathtools}
\usepackage[most]{tcolorbox}
\usepackage{hyperref}
\usepackage{lipsum}
\usepackage{empheq}
\usepackage{braket}
\usepackage{amssymb}
\usepackage{enumerate}

\usepackage{array}
\usepackage{booktabs}
\newcommand{\fnmarkdef}[1]{\footnotemark
  \expandafter\xdef\csname fn@#1\endcsname{\number\value{footnote}}}
\newcommand{\fntextdef}[2]{\footnotetext[\csname fn@#1\endcsname]{#2}}

\newcommand{\sop}{\hat{s}}

\newcommand{\bop}{\hat b}
\newcommand{\bTop}{\hat{B}}

\newcommand{\aop}{\hat a}
\newcommand{\aTop}{\hat{A}}
\newcommand{\Hop}{\hat{H}}
\newcommand{\Aop}{\hat{\mathcal{A}}}
\newcommand{\Bop}{\hat{\tilde{B}}}

\newcommand{\ain}{\hat{a}_{\scriptscriptstyle{\text{IN}}}}
\newcommand{\aout}{\hat{a}_{\scriptscriptstyle{\text{OUT}}}}

\newcommand{\kvec}{\mathbf{k}}
\newcommand{\qvec}{\mathbf{q}}
\newcommand{\rvec}{\mathbf{r}}

\newcommand{\vopvec}{\hat{\mathbf{v}}}
\newcommand{\Vopvec}{\hat{\mathbf{V}}}
\newcommand{\OUT}{\scriptscriptstyle{\text{OUT}}}
\newcommand{\IN}{\scriptscriptstyle{\text{IN}}}

\newcommand*{\rttensor}[1]{\overline{\overline{#1}}}
\newcommand{\bracket}[2]{\langle #1 | #2 \rangle}

\hypersetup{colorlinks,linkcolor=blue,citecolor=blue,urlcolor=blue}

\begin{document}

\title{
Tunable source of entangled microwaves by nonlinear scattering of magnons
}
\author{M. Brühlmann}
\affiliation{Institute for Theoretical Physics and Vienna Center for Quantum Science and Technology, TU Wien, A-1040 Vienna, Austria.}
\author{C. Gonzalez-Ballestero}
\email{carlos.gonzalez-ballestero@tuwien.ac.at}
\affiliation{Institute for Theoretical Physics and Vienna Center for Quantum Science and Technology, TU Wien, A-1040 Vienna, Austria.}

\begin{abstract}
We propose a route to prepare entangled states of microwaves propagating in spatially separated transmission lines using the intrinsic nonlinear dynamics of ferromagnetic spin waves (magnons). We first show that a driven magnon mode can produce entangled magnon pairs at low temperatures by spontaneous parametric down-conversion enabled by three-magnon processes. The anisotropic magnon properties in thin films result in strict phase-matching conditions which allow only a well-defined pair of magnon modes to be amplified. This mode selectivity allows to out-couple each amplified mode to an independent microwave transmission line, producing non-degenerate propagating microwave entanglement with current experimental parameters. In addition, we prove that the frequency difference between the two entangled microwaves can be dynamically tuned across a few-GHz range by changing the orientation of a bias magnetic field. Our figures of merit---e.g., the amount of microwave entanglement---can be largely improved by optimizing material and geometry. Our work establishes that quantum states of propagating magnons can be created by their nonlinear dynamics alone, and opens a pathway toward integrating magnons in hybrid quantum platforms.
\end{abstract}

\maketitle

Entanglement between spatially separated propagating microwave fields is a key resource for distributed continuous-variable quantum information, quantum networks, and quantum-enhanced sensing~\cite{Braunstein_Quantum_2005,Weedbrook_Gaussian_2012,ku_generating_2015,ockeloen_Noisless_2017}. It has been demonstrated using parametric processes enabled by Josephson junctions and junction-based devices~\cite{eichler_observation_2011,flurin_generating_2012,eichler_quantum_2014,Sandbo_chang_generating_2018,schneider_observation_2020,peugot_generating_2021,esposito_observation_2022,fedorov_finite-time_2018, Leppakangas_nonclassical_2013}. However, several shortcomings of Josephson junctions, such as fabrication complexity or incompatibility with magnetic fields (and thus also with certain spin qubits)~\cite{Xu_Magnetic_2023}, have motivated ongoing research into alternative physical platforms and mechanisms able to generate propagating microwave entanglement~\cite{barzanjeh_stationary_2019,mohamed_path-entangled_2024,Forgues_experimental_2015,Xu_Magnetic_2023,Wu_junction-free_2024, Di_high-efficiency_2023,yu_magnetostrictively_2020,Parker_degenerate_2022}. In this Letter, we propose a route that exploits the intrinsic nonlinearity of spin waves propagating in a ferromagnetic thin film.\par

Propagating spin waves in ferromagnets (magnons) have long been studied in the classical domain as alternative information carriers~\cite{chumak_roadmap_2022,mahmoud_introduction_2020} due to, among other properties, their strong nonlinearity stemming from 3- and 4-magnon processes~\cite{suhl_theory_1957,anderson_instability_1955,kalinikos_theory_1986,lvov_wave_1994,gurevich_magnetization_2020,boardman_three_1988,ordonez-romero_thee_2009,mohseni_controlling_2021,krivosik_hamiltonian_2010,demidov_generation_2011,chumak_magnon_2015,pirro_advances_2021}. Nonlinearity lies at the heart of core fundamental discoveries such as magnon condensates~\cite{demokritov_boseeinstein_2006} and applications such as magnon logical devices~\cite{chumak_magnon_2014,wang_magnonic_2020,bessonov_frequency-domain_2026,zenbaa_realization_2025} or neuromorphic computing~\cite{papp_nanoscale_2021,guo_integrated_2026,breitbach_all-magnonic_2026,Watt_Implementing_2021,korber_pattern_2023,Xiong_predicting_2026}. In the first part of this Letter, we show that the intrinsic nonlinearity of magnons can be exploited also in the quantum regime, specifically to prepare entangled states of propagating magnons at cryogenic temperatures by magnon-to-magnon spontaneous parametric down-conversion~\cite{boardman_three_1988,mohseni_controlling_2021,ordonez-romero_thee_2009}.  
The anisotropic properties of magnons provides strict phase-matching conditions for this process and thus enables spectral selectivity, i.e., the parametric amplification of only two specific magnon modes with well-defined frequencies instead of a broadband continuum. In addition, these frequencies can be tuned in situ through the orientation of an external magnetic field. In the second part of this work, we show that entanglement can be transferred to microwaves by coupling each of the amplified magnon modes to a microwave transmission line. 
This provides a source of spatially separated entangled microwaves with in-situ frequency selectivity. \par

\begin{figure}
    \centering
    \includegraphics[width=\linewidth]{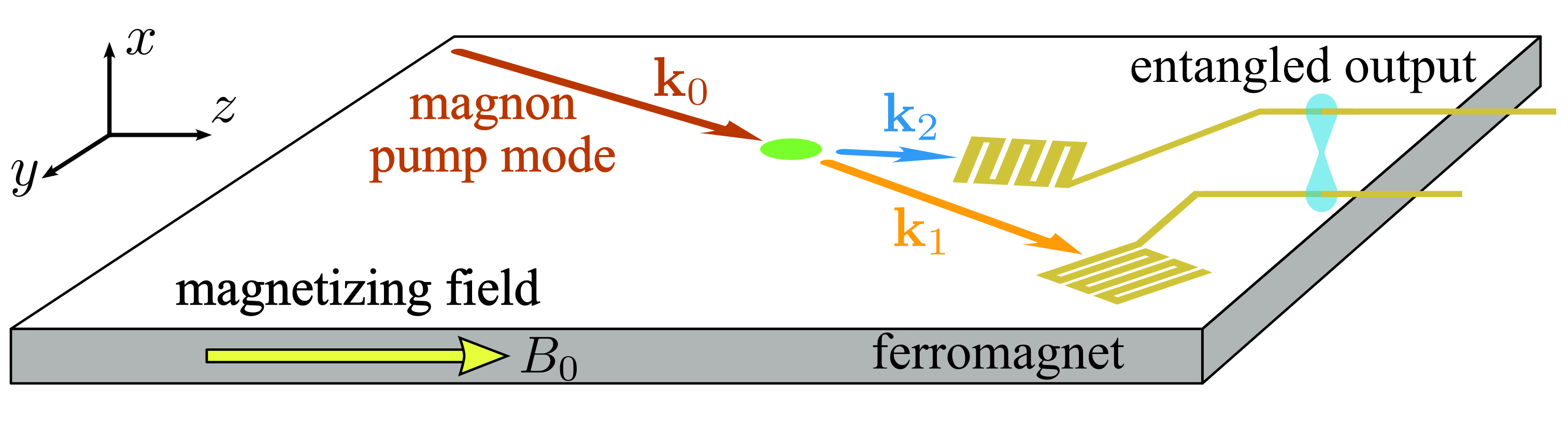}
    \caption{A magnon mode $\mathbf{k}_0$ (red) of a thin-film magnet magnetized in-plane by a field $\mathbf{B}_0$ is pumped coherently, parametrically exciting propagating modes $\mathbf{k}_1$ (orange) and $\mathbf{k}_2$ (blue) by three-magnon scattering interactions. Two antennas outcouple the amplified magnon modes to independent microwave transmission lines, resulting in entangled microwave output fields. Throughout this work we take $B_0 = 20 \,\mathrm{mT}$, film thickness $d = 40 \, \mathrm{nm}$, and parameters for yttrium iron garnet (YIG) (saturation magnetization $M_S = 1.35\times 10^5 \, \mathrm{A \, m^{-1}}$, exchange stiffness  $A_{\text{ex}} = 3.5 \, \mathrm{pJ \, m^{-1}}$, Gilbert damping coefficient $\lambda=1.1\times10^{-4}$).
    }
    \label{fig:Fig1}
\end{figure}
The system under study consists of a thin-film magnet (Fig.~\ref{fig:Fig1}) with thickness $d$ along the $x$-axis and infinite extension across the $yz$-plane. The film is magnetized along the $z$-axis by an external uniform magnetic field $\mathbf{B}_0=B_0\mathbf{e}_z$. The magnetization field can sustain propagating excitations---spin waves---labeled by a two-dimensional wavevector mode index $\mathbf{k}$ in the $yz$-plane~\cite{kalinikos_theory_1986,CGB_2022}. In the thin-film limit $kd\ll 1$~\footnote{Due to confinement along $x$, the thin-film structure supports an infinite number of bands $\omega_{n,\mathbf{k}}$ ($n=0,1,2...$). In the thin-film limit $kd\ll 1$ they are separated by large energy gaps, allowing one to consider only the $n=0$ band.}, spin waves are described by the quantum Hamiltonian
$\Hop =\hbar \int d\kvec \omega_\kvec \, \sop_\kvec^\dagger \sop_\kvec+\hat{H}_{\rm nl}$~\cite{mills_quantum2006,prabhakar2009spin,Serpico_quantum_2025,CGB_2022} with $\sop_\kvec$ bosonic operators fulfilling $[\sop_\kvec,\sop^\dagger_{\kvec'}]=\delta(\mathbf{k}-\mathbf{k}')$ and $\omega_\mathbf{k}$ the frequency of mode $\mathbf{k}$, whose expression is given in the literature~\cite{kalinikos_theory_1986,Bruehlmann_2026_classical,CGB_2022} and the Supplemental Material~\cite{supplemental_material}. The term $\hat{H}_{\rm nl}$ describes three- and four-magnon interactions arising from the intrinsic spin nonlinearity. For our chosen parameters the dominant process is three-magnon scattering~\cite{supplemental_material}, given by the momentum-conserving interaction~\cite{krivosik_hamiltonian_2010,prabhakar2009spin,Bruehlmann_2026_classical}
\begin{align}
    \Hop_\text{nl} &= \hbar \iint d\kvec\, d\kvec^\prime  \left( g(\kvec,\kvec^\prime)\sop_\kvec \sop_{\kvec^\prime} \sop_{\kvec+\kvec^\prime}^\dagger +h.c. \right). 
\end{align}
The nonlinear coupling strength $g(\kvec,\kvec^\prime)$ for thin films is given analytically in the literature and the Supplemental Material~\cite{Bruehlmann_2026_classical,supplemental_material}. Nonlinear dynamics are usually weak at low excitation numbers and therefore masked by damping and thermal fluctuations. Here we consider the opposite regime, where a single mode $\kvec_0$ is coherently driven (e.g. through a microwave antenna) strongly enough to reach the parametric instability threshold, i.e., to generate amplification of mode pairs $\{\mathbf{k},\mathbf{k}_0-\mathbf{k}\}$~\cite{melkov_parametric_1989,bagada_parametric_1997,bracher_mode_2011} in analogy to spontaneous parametric down-conversion in nonlinear optics. As shown by previous works~\cite{suhl_theory_1957,zakharov_spin-wave_1975,mohseni_parametric_2020,bracher_mode_2011,Bruehlmann_2026_classical} and as demonstrated below, a core property of magnons is that this process is mode-selective, i.e., a specific mode pair is amplified which hereafter we label $\mathbf{k}_1$ and $\mathbf{k}_2 = \mathbf{k}_0-\mathbf{k}_1$. This mode selectivity is key to our proposal, as it prevents magnon entanglement to be distributed (and thus diluted) across a wide frequency band.\par 

Since only the three modes $\{\kvec_0,\kvec_1,\kvec_2\}$ are relevant to the dynamics, we map the continuum system into a three-mode problem. We do so by projecting the dynamics onto the subspace spanned by the three collective magnon modes defined by annihilation operators $\bTop_j = \int d\kvec \, f_j(\kvec) \sop_\kvec$ ($j\in\{0,1,2\}$). The square-normalized functions $f_j(\kvec)$ are chosen as Gaussian functions centered at $\kvec_j$ whose frequency width is equal to the damping of mode $\kvec_j$, i.e., $\Delta \omega_j \sim \gamma_{\kvec_j}=(2\lambda\omega_{\kvec_j}/|\gamma_g| )(\partial\omega_{\kvec_j}/\partial B_0 )$ where $\lambda$ is the Gilbert damping coefficient and $\gamma_g$ the electron gyromagnetic ratio~\cite{supplemental_material}. Assuming that the resulting widths in wavevector space are much smaller than the separation between the center points $\kvec_j$, the discrete operators $\bTop_j$ satisfy bosonic commutation relations $[\bTop_i,\bTop_j^\dagger]=\delta_{i,j}$ and $[\bTop_i,\bTop_j]=0$. The quantum state evolution for this three-mode system is governed by the master equation
\begin{align}\label{eq:master_eq_preTransform}
    \frac{d}{dt} \hat{\rho} = -\frac{i}{\hbar} [\Hop, \hat{\rho}] + \sum_j \frac{\gamma_j}{2} \Bigl( (1+\bar{n}_j) \mathcal{L}_{\bTop_j} + \bar{n}_j \mathcal{L}_{\bTop_j^\dagger} \Bigr),
\end{align}
where $\hat{\rho}$ is the density matrix and with the Hamiltonian
\begin{align}\label{eq:lab_frame_Hamiltonian}
    \frac{\Hop}{\hbar} = \sum_{j=0}^2 \omega_{\kvec_j} \bTop_j^\dagger \bTop_j &+ (G \, \bTop_1 \bTop_2 \bTop_0^\dagger + i \alpha_\text{d} e^{-i\omega_{\kvec_0} t} \bTop_0^\dagger + h.c. ).
\end{align}
Here, we have included the coherent resonant drive of mode $\bTop_0$ with rate $\alpha_d$ and defined the nonlinear coupling rate $G$, whose analytical expression is given in the Supplemental Material~\cite{supplemental_material}. The second term in Eq.~\eqref{eq:master_eq_preTransform} describes the magnon loss due to Gilbert damping, with $\gamma_j=\gamma_{\kvec_j}$ the damping rate of mode $j$~\cite{CGB_2022,supplemental_material}, $\bar{n}_j$ the Bose-Einstein distribution at frequency $\omega_{\mathbf{k}_j}$, and with the Lindblad operators $\mathcal{L}_{\bTop_j} = \bTop_j \hat{\rho} \bTop_j^\dagger - \{\bTop_j^\dagger \bTop_j , \hat{\rho}\}/2$.
\par

\begin{figure}[t] 
    \centering
    \includegraphics[width=\linewidth]{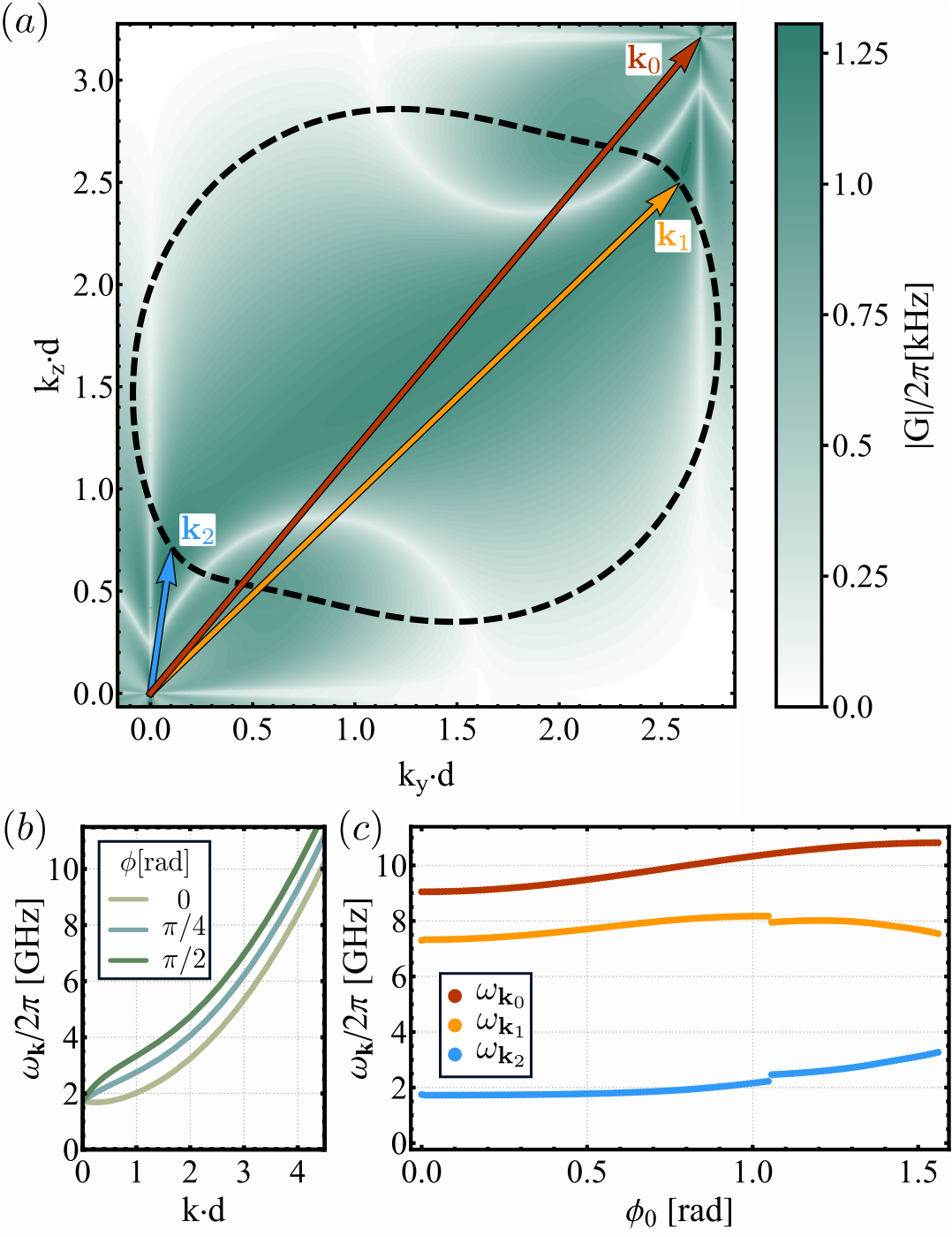}
    \caption{
   Mode selection in parametric excitation $\mathbf{k}_0\to\{\mathbf{k}_1,\mathbf{k}_2\}$ with $\mathbf{k}_2=\mathbf{k}_0-\mathbf{k}_1$. (a) Three-magnon coupling rate $|G|$ as a function of $\mathbf{k}_1$, for a driven mode with $|\kvec_0|=2\pi/60\,\mathrm{nm^{-1}}$ and propagation angle  $\phi_0=40\,\deg$ (red arrow) to the applied field $\mathbf{B}_0$. The dashed line marks the subset of $\mathbf{k}_1$ for which energy is conserved in the process, $\omega_{\kvec_0}=\omega_{\kvec_1}+ \omega_{\kvec_2}$. Within this subset, only the modes maximizing $\vert G\vert$ (blue and orange arrows) are parametrically amplified.
    (b) Magnon dispersion relation $\omega_\kvec$ at different propagation directions characterized by the angle $\phi$ between $\mathbf{B}_0$ and $\kvec$.
    (c) Frequencies of the driven (red) and parametrically amplified modes (orange and blue) as a function of $\phi_0$.
    }
    \label{fig:mode_tunability}
\end{figure} 
We first describe the classical nonlinear dynamics of the three-mode system. We define the slowly varying variables $\beta_j \equiv\langle \bTop_j\rangle e^{i\omega_{\mathbf{k}_j} t}$ corresponding to the magnon amplitudes in a rotating frame~\footnote{We implicitly assume in this definition that the three-magnon scattering conserves energy, $\omega_{\kvec_0}=\omega_{\kvec_1}+\omega_{\kvec_2}$, and that magnon modes fulfilling such resonance condition exist in the film, which is true for the parameters of this work. If this is not the case our treatment can easily be generalized~\cite{Bruehlmann_2026_classical}.}. This frame is a convenient choice to eliminate explicit time dependence from the equations of motion. We then derive from Eq.~\eqref{eq:master_eq_preTransform} the equations of motion for $\beta_j$ and take the classical limit by neglecting all correlations, e.g.  $\langle \bTop_j \bTop_k\rangle \approx \beta_j \beta_k e^{-i(\omega_{\kvec_j}+\omega_{\kvec_k})t}$ (mean-field approximation~\cite{Bruehlmann_2026_classical,Kubo_generalized_1962,Kadanoff_meanfield_2009}). This results in the nonlinear equations of motion
\begin{align}
    \dot{\beta}_0 &= -(\gamma_0/2) \beta_0 -i G \,\beta_1 \beta_2 +\alpha_d ,\\
    \dot{\beta}_1 &= -(\gamma_1/2) \beta_1 -i G^* \, \beta_2^* \beta_0 ,\\
    \dot{\beta}_2 &= -(\gamma_2/2) \beta_2 -i G^* \, \beta_1^* \beta_0.
\end{align}
Their steady-state solution has two qualitatively different regimes. Below the parametric instability threshold  $|\alpha_\text{d}|\leq|\alpha_\text{th}|$, with
\begin{equation}\label{eq:threshold}
    |\alpha_{\text{th}}| = \gamma_0\sqrt{\gamma_1\gamma_2} /(4\vert G\vert),
\end{equation}
the solution is $\beta_{1,2}=0$ and $\beta_0\propto\alpha_d$. We focus on the regime above the threshold, where modes $\kvec_1$ and $\kvec_2$ become parametrically amplified with amplitudes~\cite{Bruehlmann_2026_classical,lee_nonlinear_2023}
\begin{equation}
    |\beta_1|^2 = \frac{\gamma_2}{\gamma_1}|\beta_2|^2
    = \frac{\gamma_0 \gamma_2}{4 |G|^2} \left( \frac{|\alpha_{\rm d}|}{|\alpha_{\rm th}|}-1 \right),
\end{equation}
while the driven mode amplitude stays constant, $\beta_0=2|\alpha_{\rm th}| e^{i\mathrm{arg}(\alpha_{\rm d})}/\gamma_0$.
As we will see below, in this regime the fluctuations of the amplified modes become entangled.\par

Two key constraints determine which modes, i.e., which specific values of wavevectors $\mathbf{k}_{1,2}$, are amplified in this process. The first is conservation of momentum and---in low-loss magnets such as YIG ($\omega_{\kvec_j}/\gamma_j\gg1$)---also energy, i.e., $\omega_{\kvec_0}=\omega_{\kvec_1}+\omega_{\kvec_2}$~\cite{Bruehlmann_2026_classical}. This condition, however, is still fulfilled by a continuum of modes, as shown in Fig.~\ref{fig:mode_tunability}(a). Here, the dashed line indicates all modes $\mathbf{k}_1$ for which energy and momentum are conserved, for a specific driven mode $\kvec_0$ (red arrow). The second is the fact that the threshold drive for parametric instability, Eq.~\eqref{eq:threshold}, depends on the three-magnon coupling $G(\kvec_0,\kvec_1,\kvec_2)$ which, for magnons, is strongly wavevector-dependent as shown by the density-plot background of Fig.~\ref{fig:mode_tunability}(a). Thus, as driving strength $\alpha_d$ increases, the threshold will be reached first by the more strongly coupled pair of modes (orange and blue arrows in the figure) while any other mode remains below threshold~\cite{suhl_theory_1957,zakharov_spin-wave_1975,Bruehlmann_2026_classical,Hwang_2026_Harmonic}. Increasing the drive strength further will not excite new modes as the amplitude $\beta_0$ remains constant. Importantly, the frequencies of the two parametrically amplified modes can be tuned by modifying the angle $\phi_0$ between the directions of the driven mode $\mathbf{k}_0$ and the applied magnetic field $\mathbf{B}_0$. 
This is because the magnon dispersion relation, and with it the nonlinear coupling rates and threshold conditions, depend on its propagation direction (see Fig.~\ref{fig:mode_tunability}(b)). As shown in Fig.~\ref{fig:mode_tunability}(c) the angle $\phi_0$ provides an external control to tune the frequency difference between the outgoing modes across a few-GHz range. \par

\begin{figure}[t]
    \centering
    \includegraphics[width=\linewidth]{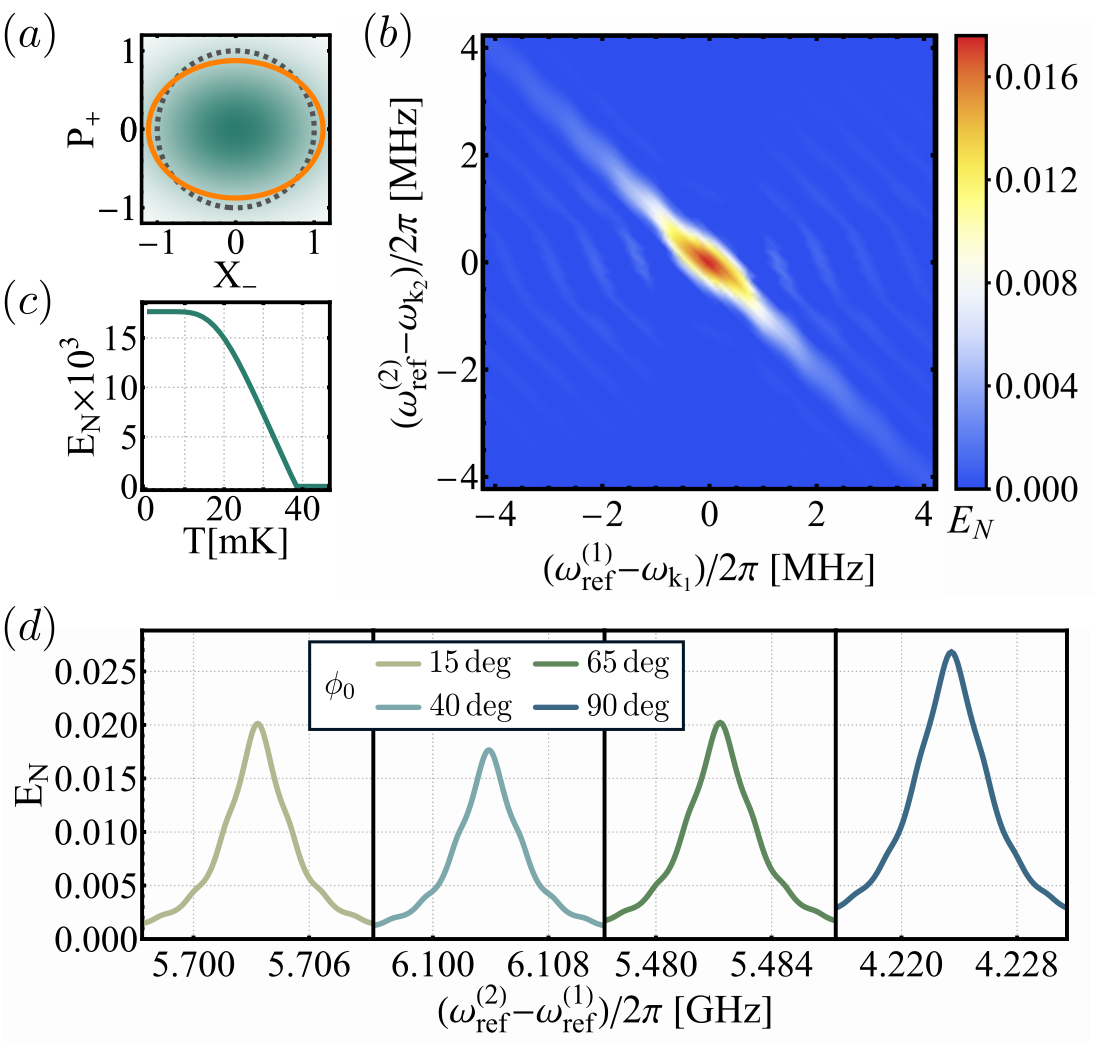}
    \caption{ Entanglement for $|\kvec_0|=2\pi/60\,\mathrm{nm^{-1}}$, $\phi_0 = 40\,\deg$, $|\kappa_{j}|=122.5\,\mathrm{kHz}$, and $\alpha_\text{d}=1.2 \alpha_\text{th}$. (a) Projection of the two-mode magnon Wigner function onto the EPR-quadrature plane, showing two-mode squeezing. (b) Entanglement, quantified by logarithmic negativity as a function of reference frequencies, between temporal modes (Eq.~\ref{eq:heterodyne_operators}) of each transmission line. (c) Temperature dependence of the entanglement peak in panel (b). (d) Spectral selectivity of microwave entanglement: logarithmic negativity along the energy conservation line $\omega_\text{ref}^{(1)}+\omega_\text{ref}^{(2)}=\omega_{\kvec_0}$ as a function of temporal mode detuning, for different propagation angles $\phi_0$ of the driven magnon mode $\mathbf{k}_0$. 
    }
    \label{fig:entanglement}
\end{figure}

If temperature and dissipation are low, the fluctuations of the outgoing modes $\mathbf{k}_1$ and $\mathbf{k}_2$ are entangled, in analogy to photons produced in down-conversion processes~\cite{reid_quantum_1988,reid_demonstration_1989,ou_realization_1992}. To certify this, we write the mode operators as $\bTop_j(t) = e^{i\Omega_j t} ( \beta_j + \bop_j(t) )$, where $\bop_j$ describe the fluctuations of each mode above the classical amplitude $\beta_j$. Introducing this decomposition together with the steady-state expressions for $\beta_j$ into the master equation~\eqref{eq:master_eq_preTransform} we obtain a similar master equation for the fluctuations, with a modified Hamiltonian \begin{align}\label{eq:eff_lin_int_Hamiltonian}
    \Hop_f = \hbar G\, \bigl(\beta_0^* \bop_1 \bop_2 +\bop_0^\dagger (\beta_1 \bop_2 + \beta_2 \bop_1) \bigr)+ h.c. 
\end{align}
Here, we have neglected a nonlinear term $\Hop_\text{f,nl} = \hbar G \, \bop_1 \bop_2 \bop_0^\dagger +h.c. $ assuming $|\beta_j|\gg1$ (linearization), which is fulfilled by our choice of parameters. The two-mode squeezing term $\propto \hat{b}_1\hat{b}_2$ in Eq.~\eqref{eq:eff_lin_int_Hamiltonian} describes the creation of correlated magnon pairs by the classical drive, and is the responsible for entanglement generation~\cite{mycroftquantum_2026}. To quantify entanglement between modes $\hat b_1$ and $\hat b_2$ we define the EPR quadratures $\hat{X}_\pm = (\hat{x}_1 \pm \hat x_2\cos\varphi + \hat p_2\sin\varphi)/\sqrt{2}$ and $\hat{P}_\pm=(\hat{p}_1 \pm \hat p_2\cos\varphi - \hat x_2\sin\varphi)/\sqrt{2}$, where $\hat{x}_j=( \bop_j^\dagger +\bop_j) /\sqrt{2}$, $\hat{p}_j= i (\bop_j^\dagger -\bop_j)/\sqrt{2}$, and $\varphi = \arg\langle \bop_1 \bop_2\rangle$, and analytically compute their steady-state variances from the master equation. The Duan-Simon criterion for entanglement $\text{Var}(\hat{X}_-)+\text{Var}(\hat{P}_+)<1$~\cite{duan_inseparability_2000,simon_peres-horodecki_2000} is clearly fulfilled as shown by Fig~\ref{fig:entanglement}(a). This provides the first proposal for nonclassical state preparation of propagating magnons using their nonlinar dynamics alone. \par

The final result of this work is the possibility to transfer the frequency-selective entanglement of magnons into microwaves. Specifically, we propose to couple two microwave transmission lines to the magnons via inductive antennas, whose coupling to propagating magnons can be spectrally engineered through geometry~\cite{sethares_magnetostatic_1979,connelly_efficient_2021,hofinger_k-selective_2026,casulleras_generation_2023}. As shown in~\cite{supplemental_material}, this allows one to couple independently each transmission line $j\in\{1,2\}$ to one of the outgoing modes $\hat b_1$ and $\hat b_2$. To describe the full magnon-transmission line system we add to the master equation~\eqref{eq:master_eq_preTransform} the free Hamiltonian of the transmission lines, $\Hop_{\rm mw} = \sum_{j=1}^2\int_0^\infty d\omega\, \hbar \omega \aTop_\omega^{(j)\dagger} \aTop_\omega^{(j)}$, and the microwave-magnon interaction Hamiltonian, $\Hop_{\text{int}} = i\hbar \sum_{j=1}^2 \int d\omega\, \sqrt{\kappa_{j}} \aTop_\omega^{(j)} \bTop_j^\dagger + h.c. $ where $\aTop_\omega^{(j)}$ are  ladder operators for microwave modes at frequency $\omega$ obeying bosonic commutation relations, $[\aTop_\omega^{(j)}, \aTop_{\omega^\prime}^{(k)\dagger}]=\delta_{j,k}\delta(\omega-\omega^\prime)$ and $[\aTop_\omega^{(j)}, \aTop_{\omega^\prime}^{(k)}]=0$. The magnon-photon coupling rate is assumed weak enough to negligibly affect the magnon dynamics, i.e., $2\pi|\kappa_{j}| \ll \gamma_j$. We follow a similar procedure as above, i.e., we split magnon and microwave operators into classical part and fluctuations in a rotating frame, $\bTop_j(t) = e^{-i\omega_{\mathbf{k}_j} t} ( \beta_j + \bop_j(t) )$ and $\aTop_\omega^{(j)}(t) = e^{-i\omega_{\mathbf{k}_j} t} ( \alpha_j(\omega) + \aop_\omega^{(j)}(t) )$, and analytically compute the mean field amplitudes $\beta_j$ and $\alpha_j(\omega)$ in the long-time limit $\kappa_{j}\,t\gg 1$ (see Supplemental Material~\cite{supplemental_material}). Using this solution we linearize the dynamics to obtain a linear system of equations for the fluctuations $\bop_j$ and $\aop_\omega^{(j)}$. Being interested in state transfer, we derive for the latter the input-output relation~\cite{gardiner_quantum_2004}
\begin{align}\label{eq:input_output}
    \aout^{(j)}(t) = \ain^{(j)}(t) -\sqrt{2\pi\kappa_{j}} \bop_j(t).
\end{align}
Here, the microwave input and output operators describe the freely-propagating transmission line modes in the distant past and the distant future. They are defined as $\ain^{(j)} (t) = (2\pi)^{-1/2}\lim_{t_0\to-\infty} \int d\omega\, e^{-i\Delta_\omega^{(j)}(t-t_0)} \aop_\omega^{(j)}(t_0)$ with $\Delta_\omega^{(j)}= \omega-\omega_{\mathbf{k}_j}$;  $\aout^{(j)} (t)$ has identical definition under the substitution $-\infty\leftrightarrow\infty$. These operators obey the temporal commutation relations $[\ain^{(j)}(t),\ain^{(j^\prime)\dagger}(t^\prime)]= \delta_{j,j^\prime} \delta(t-t^\prime)$ and $[\ain^{(j)}(t),\aout^{(j^\prime)\dagger}(t^\prime)]= 0$. \par

The input-output relation suggests that the modes $\aop_\omega^{(j)}(t)$, i.e., the fluctuations of the microwave field in a rotating frame, can be entangled when the magnon modes $\bop_1$ and $\bop_2$ are. Experimentally, these fluctuations can be measured by an interferometric heterodyne detection scheme~\cite{walls_quantum_2008,eichler_observation_2011,morin_experimentally_2013}. Thus, to quantify the microwave entanglement, we define the temporal mode bosonic ladder operators describing the detected output, which in the original frame of Eqs~\eqref{eq:master_eq_preTransform}-\eqref{eq:lab_frame_Hamiltonian} read
\begin{align}\label{eq:heterodyne_operators}
    \Aop_j\left(\omega_\text{ref}^{(j)}\right) = \int dt\, F_j(t) e^{-i\omega_\text{ref}^{(j)} t} \, \aout^{(j)} (t) \, e^{-i\omega_{\kvec_j}t}  .
\end{align}
Here, $\omega_\text{ref}^{(j)}$ is the frequency of the reference signal and the square-normalized function $F_j(t)$ is the temporal filter function of the detector. The oscillating exponential with frequency $\omega_{\kvec_j}$ arises from the transformation to the original frame. Using the input-output relation of Eq.~\eqref{eq:input_output} and the magnon dynamical equations derived above, we derive analytically the covariance matrix of the two temporal modes $\Aop_j(\omega_\text{ref}^{(j)})$ in the long-time limit~\cite{supplemental_material}. From this, we quantify the microwave entanglement by the logarithmic negativity $E_N=\max(0,-\ln(2\tilde{\nu}_-))$, with $\tilde{\nu}_-$ the smallest symplectic eigenvalue of the partially transposed covariance matrix~\cite{duan_inseparability_2000,simon_peres-horodecki_2000}. The Duan-Simon criterion for entanglement is $\tilde{\nu}_-<1/2$ or equivalently $E_N>0$.\par

Figure~\ref{fig:entanglement}(b) shows the logarithmic negativity $E_N$ of the two temporal modes as a function of their reference frequencies $\omega_{\text{ref}}^{(1,2)}$ at $T\lesssim 10\,\mathrm{mK}$. We choose both $F_j(t)$ to be rectangular step functions with a large width (here, $\Delta T=120\,\mathrm{\mu s}$) to spectrally filter out most of the thermal background. 
As Figure~\ref{fig:entanglement}(b) shows, entanglement between the microwave modes is significant when their reference frequencies match the frequencies of the parametrically amplified magnons within the magnons' natural linewidth ($\vert\omega_\text{ref}^{(j)}-\omega_{\kvec_j}\vert \lesssim \gamma_j$ for $j=1,2$). Away from this energy-conservation region, the terms $\exp(-i\omega^{(j)}_{\rm ref}t)$ do not cancel out effectively the oscillatory time dependence of the magnon two-time correlators, and entanglement is averaged out during the filtering time $\Delta T$. In Figure~\ref{fig:entanglement}(c) we show how the temperature affects the maximum entanglement observed in Fig.~\ref{fig:entanglement}(b), confirming the existence of entanglement up to several tens of $\rm mK$, consistent with the regime of low thermal occupations $\bar{n}_j$. Finally, Fig.~\ref{fig:entanglement}(d) shows the microwave logarithmic negativity along the energy conservation line for different angles $\phi_0$ between the driven magnon mode $\mathbf{k}_0$ and the applied field $\mathbf{B}_0$. The frequency difference between the maximally entangled modes follows exactly that of the amplified magnons, shown in Figure~\ref{fig:mode_tunability}(c). Our results unequivocally confirm that microwave mode entanglement is a consequence of entanglement between propagating magnons, and therefore that the angle $\phi_0$, a common control knob in magnonics~\cite{Qin_propagating_2018,jiang_thin-film_2026,lim_tracking_2020}, can be used to distribute the microwave entanglement across frequency space in a controlled way. In other words, our results show that magnons enable to prepare two-mode entangled microwave states and tune their frequency detuning across a $\sim 2$-GHz bandwidth.\par

To conclude, we propose three-magnon scattering as a new mechanism for preparing and distributing microwave photon entanglement with frequency selectivity. The mode selectivity of nonlinear magnon scattering enables the use of simple geometries (in this case, a thin film) without the need for additional cavities or other external phase-matching mechanisms. In addition, magnons offer an accessible control knob to tune the frequency of the entangled photons, namely the direction of the applied magnetic field $\mathbf{B}_0$. Our proof-of-principle results are just the first step: figures of merit such as the degree of entanglement or tunability range can be improved through simple extensions of our model to other materials, field orientations, and geometries, or by exploiting additional control knobs such as the magnitude of $\mathbf{B}_0$. Besides the spectral mode selectivity exploited in this work, the directionality of the entangled magnon modes is also controllable~\cite{supplemental_material}, potentially enabling entanglement distribution in more complex multi-channel devices beyond only two transmission lines. Our work establishes the first route to preparing nonclassical states of propagating magnons, advancing a core goal of the quantum magnonics community~\cite{chumak_roadmap_2022,li_strong_2026,jiang_integrating_2023,li_hybrid_2020,serha_ultralong-living_2026,schmoll_brillouin_2026,serha_magnetic_2024}. Furthermore, the flexibility of the proposed mechanism, which can be applied to other nonlinear processes such as nonlinear magnon-photon~\cite{Elyasi_resources_2020,Meese_unpublished,zare_rameshti_cavity_2022,kani_squeezed_2025,safonov_nonlinear_1996,liang-sheng_photonmagnon_2005,qing-fan_nonlinear_2003} or magnon-phonon scattering~\cite{Bruehlmann_2026_classical,Hwang_2026_Harmonic,rivard_parametric_2025,rivard_sustained_2025,jander_parametric_2025,Chowdhury_nondegenerate_2017,zhang_cavity_2016,Bittencourt_Magnomechanical_2023,Cherkasskii_Theory_2025,geilen_parametric_2025,mohseni_amplification_2023,lisenkov_magnetoelastic_2019,rückriegel_magnetoelastic_2014}, paves the way toward integration of magnons in microwave hybrid quantum platforms~\cite{lachance-quirion_hybrid_2019,Li_Hybrid-Magnon_2022,Awschalom_quantum_2021,wang_room-temperature_2026}.\\

This research was funded in part by the Austrian Science Fund (FWF) [10.55776/PIN3404324]

\clearpage
\medskip 
\bibliographystyle{apsrev4-2}
\bibliography{bibliography}

@misc{supplemental_material,
        note = {{S}ee supplemental material for the thin-film magnon mode structure and rates, the derivation of the dynamics in the collective-mode basis, the description of coupling to transmission lines and the mean-field solutions of the coupled system, and the calculation of entanglement of the heterodyne temporal modes.}
}

@article{CGB_2022,
  title = {Towards a quantum interface between spin waves and paramagnetic spin baths},
  author = {Gonzalez-Ballestero, C. and van der Sar, Toeno and Romero-Isart, O.},
  journal = {Phys. Rev. B},
  volume = {105},
  issue = {7},
  pages = {075410},
  numpages = {43},
  year = {2022},
  month = {Feb},
  publisher = {American Physical Society},
  doi = {10.1103/PhysRevB.105.075410},
  url = {https://link.aps.org/doi/10.1103/PhysRevB.105.075410}
}

@article{gonzalez-ballestero_quantum_2020,
	title = {Quantum Acoustomechanics with a Micromagnet},
	volume = {124},
	url = {https://link.aps.org/doi/10.1103/PhysRevLett.124.093602},
	doi = {10.1103/PhysRevLett.124.093602},
	pages = {093602},
	number = {9},
	journaltitle = {Physical Review Letters},
	journal = {Phys. Rev. Lett.},
	author = {Gonzalez-Ballestero, Carlos and Gieseler, Jan and Romero-Isart, Oriol},
	urldate = {2023-06-16},
	date = {2020-03-03},
        year={2020}
}

@article{gonzalez-ballestero_theory_2020,
	title = {Theory of quantum acoustomagnonics and acoustomechanics with a micromagnet},
	volume = {101},
	issn = {2469-9950, 2469-9969},
	url = {https://link.aps.org/doi/10.1103/PhysRevB.101.125404},
	doi = {10.1103/PhysRevB.101.125404},
	pages = {125404},
	number = {12},
        journal = {Phys. Rev. B},
	journaltitle = {Physical Review B},
	shortjournal = {Phys. Rev. B},year={2020},
	author = {Gonzalez-Ballestero, C. and Hümmer, D. and Gieseler, J. and Romero-Isart, O.},
	urldate = {2023-01-12},
	date = {2020-03-03}
}

@article{Bruehlmann_2026_classical,
  title = {Classical and quantum theory of magnonic and magnetoelastic nonlinear dynamics in continuum geometries},
  author = {Br\"uhlmann, Marco and Hwang, Yunyoung and Puebla, Jorge and Gonzalez-Ballestero, Carlos},
  journal = {Phys. Rev. B},
  volume = {113},
  issue = {1},
  pages = {014430},
  numpages = {24},
  year = {2026},
  month = {Jan},
  publisher = {American Physical Society},
  doi = {10.1103/d12l-krcy},
  url = {https://link.aps.org/doi/10.1103/d12l-krcy}
}

@article{Hwang_2026_Harmonic,
  title = {Harmonic and subharmonic magnon generation in a surface-acoustic-wave resonator},
  author = {Hwang, Yunyoung and Liao, Liyang and Puebla, Jorge and Br\"uhlmann, Marco and Gonzalez-Ballestero, Carlos and Kondou, Kouta and Ogawa, Naoki and Maekawa, Sadamichi and Otani, Yoshichika},
  journal = {Phys. Rev. Appl.},
  volume = {25},
  issue = {3},
  pages = {034056},
  numpages = {7},
  year = {2026},
  month = {Mar},
  publisher = {American Physical Society},
  doi = {10.1103/mm21-ctsb},
  url = {https://link.aps.org/doi/10.1103/mm21-ctsb}
}

@article{Celeghini_2021_Hermite,
	title = {Hermite {Functions} and {Fourier} {Series}},
	volume = {13},
	copyright = {https://creativecommons.org/licenses/by/3.0/},
	issn = {2073-8994},
	url = {https://www.mdpi.com/2073-8994/13/5/853},
	doi = {10.3390/sym13050853},
	language = {en},
	number = {5},
	urldate = {2026-11-16},
	journal = {Symmetry},
	publisher = {Multidisciplinary Digital Publishing Institute},
	author = {Celeghini, Enrico and Gadella, Manuel and del Olmo, Mariano A.},
	month = may,
	year = {2021},
	pages = {853},
}

@article{suhl_theory_1957,
	title = {The theory of ferromagnetic resonance at high signal powers},
	volume = {1},
	issn = {0022-3697},
	url = {https://www.sciencedirect.com/science/article/pii/0022369757900100},
	doi = {10.1016/0022-3697(57)90010-0},
	number = {4},
	urldate = {2026-08-28},
	journal = {J. Phys. Chem. Solids},
	author = {Suhl, H.},
	month = jan,
	year = {1957},
	pages = {209--227}
}

@article{ou_realization_1992,
	title = {Realization of the {Einstein}-{Podolsky}-{Rosen} paradox for continuous variables},
	volume = {68},
	url = {https://link.aps.org/doi/10.1103/PhysRevLett.68.3663},
	doi = {10.1103/PhysRevLett.68.3663},
	number = {25},
	urldate = {2025-09-11},
	journal = {Phys. Rev. Lett.},
	publisher = {American Physical Society},
	author = {Ou, Z. Y. and Pereira, S. F. and Kimble, H. J. and Peng, K. C.},
	month = jun,
	year = {1992},
	pages = {3663--3666},
}

@article{reid_demonstration_1989,
	title = {Demonstration of the {Einstein}-{Podolsky}-{Rosen} paradox using nondegenerate parametric amplification},
	volume = {40},
	url = {https://link.aps.org/doi/10.1103/PhysRevA.40.913},
	doi = {10.1103/PhysRevA.40.913},
	number = {2},
	urldate = {2025-09-11},
	journal = {Phys. Rev. A},
	publisher = {American Physical Society},
	author = {Reid, M. D.},
	month = jul,
	year = {1989},
	pages = {913--923},
}

@article{reid_quantum_1988,
	title = {Quantum {Correlations} of {Phase} in {Nondegenerate} {Parametric} {Oscillation}},
	volume = {60},
	url = {https://link.aps.org/doi/10.1103/PhysRevLett.60.2731},
	doi = {10.1103/PhysRevLett.60.2731},
	number = {26},
	urldate = {2025-09-11},
	journal = {Phys. Rev. Lett.},
	publisher = {American Physical Society},
	author = {Reid, M. D. and Drummond, P. D.},
	month = jun,
	year = {1988},
	pages = {2731--2733},
}

@article{lachance-quirion_hybrid_2019,
	title = {Hybrid quantum systems based on magnonics},
	volume = {12},
	issn = {1882-0786},
	url = {https://dx.doi.org/10.7567/1882-0786/ab248d},
	doi = {10.7567/1882-0786/ab248d},
	language = {english},
	number = {7},
	urldate = {2024-09-02},
	journal = {Appl. Phys. Express},
	publisher = {IOP Publishing},
	author = {Lachance-Quirion, Dany and Tabuchi, Yutaka and Gloppe, Arnaud and Usami, Koji and Nakamura, Yasunobu},
	month = jun,
	year = {2019},
	pages = {070101},
}

@article{lee_nonlinear_2023,
	title = {Nonlinear {Magnon} {Polaritons}},
	volume = {130},
	url = {https://link.aps.org/doi/10.1103/PhysRevLett.130.046703},
	doi = {10.1103/PhysRevLett.130.046703},
	number = {4},
	urldate = {2024-09-11},
	journal = {Phys. Rev. Lett.},
	publisher = {American Physical Society},
	author = {Lee, Oscar and Yamamoto, Kei and Umeda, Maki and Zollitsch, Christoph W. and Elyasi, Mehrdad and Kikkawa, Takashi and Saitoh, Eiji and Bauer, Gerrit E. W. and Kurebayashi, Hidekazu},
	month = jan,
	year = {2023},
	pages = {046703},
}

@article{Elyasi_resources_2020,
  title = {Resources of nonlinear cavity magnonics for quantum information},
  author = {Elyasi, Mehrdad and Blanter, Yaroslav M. and Bauer, Gerrit E. W.},
  journal = {Phys. Rev. B},
  volume = {101},
  issue = {5},
  pages = {054402},
  numpages = {23},
  year = {2020},
  month = {Feb},
  publisher = {American Physical Society},
  doi = {10.1103/PhysRevB.101.054402},
  url = {https://link.aps.org/doi/10.1103/PhysRevB.101.054402}
}

@article{flurin_generating_2012,
  title = {Generating Entangled Microwave Radiation Over Two Transmission Lines},
  author = {Flurin, E. and Roch, N. and Mallet, F. and Devoret, M. H. and Huard, B.},
  journal = {Phys. Rev. Lett.},
  volume = {109},
  issue = {18},
  pages = {183901},
  numpages = {5},
  year = {2012},
  month = {Oct},
  publisher = {American Physical Society},
  doi = {10.1103/PhysRevLett.109.183901},
  url = {https://link.aps.org/doi/10.1103/PhysRevLett.109.183901}
}

@article{Sandbo_chang_generating_2018,
  title = {Generating Multimode Entangled Microwaves with a Superconducting Parametric Cavity},
  author = {Sandbo Chang, C. W. and Simoen, M. and Aumentado, Jos\'e and Sab\'{\i}n, Carlos and Forn-D\'{\i}az, P. and Vadiraj, A. M. and Quijandr\'{\i}a, Fernando and Johansson, G. and Fuentes, I. and Wilson, C. M.},
  journal = {Phys. Rev. Appl.},
  volume = {10},
  issue = {4},
  pages = {044019},
  numpages = {8},
  year = {2018},
  month = {Oct},
  publisher = {American Physical Society},
  doi = {10.1103/PhysRevApplied.10.044019},
  url = {https://link.aps.org/doi/10.1103/PhysRevApplied.10.044019}
}

@article{esposito_observation_2022,
  title = {Observation of Two-Mode Squeezing in a Traveling Wave Parametric Amplifier},
  author = {Esposito, Martina and Ranadive, Arpit and Planat, Luca and Leger, S\'ebastien and Fraudet, Dorian and Jouanny, Vincent and Buisson, Olivier and Guichard, Wiebke and Naud, C\'ecile and Aumentado, Jos\'e and Lecocq, Florent and Roch, Nicolas},
  journal = {Phys. Rev. Lett.},
  volume = {128},
  issue = {15},
  pages = {153603},
  numpages = {7},
  year = {2022},
  month = {Apr},
  publisher = {American Physical Society},
  doi = {10.1103/PhysRevLett.128.153603},
  url = {https://link.aps.org/doi/10.1103/PhysRevLett.128.153603}
}

@article{schneider_observation_2020,
  title = {Observation of Broadband Entanglement in Microwave Radiation from a Single Time-Varying Boundary Condition},
  author = {Schneider, B. H. and Bengtsson, A. and Svensson, I. M. and Aref, T. and Johansson, G. and Bylander, Jonas and Delsing, P.},
  journal = {Phys. Rev. Lett.},
  volume = {124},
  issue = {14},
  pages = {140503},
  numpages = {5},
  year = {2020},
  month = {Apr},
  publisher = {American Physical Society},
  doi = {10.1103/PhysRevLett.124.140503},
  url = {https://link.aps.org/doi/10.1103/PhysRevLett.124.140503}
}

@article{connelly_efficient_2021,
	title = {Efficient electromagnetic transducers for spin-wave devices},
	volume = {11},
	copyright = {2021 The Author(s)},
	issn = {2045-2322},
	url = {https://www.nature.com/articles/s41598-021-97627-3},
	doi = {10.1038/s41598-021-97627-3},
	language = {english},
	number = {1},
	urldate = {2026-09-07},
	journal = {Sci. Rep.},
	publisher = {Nature Publishing Group},
	author = {Connelly, David A. and Csaba, Gyorgy and Aquino, Hadrian Renaldo O. and Bernstein, Gary H. and Orlov, Alexei and Porod, Wolfgang and Chisum, Jonathan},
	month = sep,
	year = {2021},
	pages = {18378},
}

@ARTICLE{sethares_magnetostatic_1979,
  author={Sethares, J.C.},
  journal={IEEE Trans. Microwave Theory Tech.}, 
  title={Magnetostatic Surface-Wave Transducers}, 
  year={1979},
  volume={27},
  number={11},
  pages={902-909},
  doi={10.1109/TMTT.1979.1129760}}

@article{zare_rameshti_cavity_2022,
	title = {Cavity magnonics},
	volume = {979},
	issn = {0370-1573},
	url = {https://www.sciencedirect.com/science/article/pii/S0370157322002460},
	doi = {10.1016/j.physrep.2022.06.001},
	urldate = {2026-09-08},
	journal = {Phys. Rep.},
	author = {Zare Rameshti, Babak and Viola Kusminskiy, Silvia and Haigh, James A. and Usami, Koji and Lachance-Quirion, Dany and Nakamura, Yasunobu and Hu, Can-Ming and Tang, Hong X. and Bauer, Gerrit E. W. and Blanter, Yaroslav M.},
	month = sep,
	year = {2022},
	pages = {1--61},
}

@article{yu_magnetostrictively_2020,
  title = {Magnetostrictively Induced Stationary Entanglement between Two Microwave Fields},
  author = {Yu, Mei and Shen, Heng and Li, Jie},
  journal = {Phys. Rev. Lett.},
  volume = {124},
  issue = {21},
  pages = {213604},
  numpages = {6},
  year = {2020},
  month = {May},
  publisher = {American Physical Society},
  doi = {10.1103/PhysRevLett.124.213604},
  url = {https://link.aps.org/doi/10.1103/PhysRevLett.124.213604}
}

@article{kalinikos_theory_1986,
	title = {Theory of dipole-exchange spin wave spectrum for ferromagnetic films with mixed exchange boundary conditions},
	volume = {19},
	issn = {0022-3719},
	url = {https://dx.doi.org/10.1088/0022-3719/19/35/014},
	doi = {10.1088/0022-3719/19/35/014},
	language = {english},
	number = {35},
	urldate = {2023-12-06},
	journal = {J. Phys. C: Solid State Phys.},
	author = {Kalinikos, B. A. and Slavin, A. N.},
	month = dec,
	year = {1986},
	pages = {7013}
}

@article{Braunstein_Quantum_2005,
  title = {Quantum information with continuous variables},
  author = {Braunstein, Samuel L. and van Loock, Peter},
  journal = {Rev. Mod. Phys.},
  volume = {77},
  issue = {2},
  pages = {513--577},
  numpages = {0},
  year = {2005},
  month = {Jun},
  publisher = {American Physical Society},
  doi = {10.1103/RevModPhys.77.513},
  url = {https://link.aps.org/doi/10.1103/RevModPhys.77.513}
}

@article{Weedbrook_Gaussian_2012,
  title = {Gaussian quantum information},
  author = {Weedbrook, Christian and Pirandola, Stefano and Garc\'{\i}a-Patr\'on, Ra\'ul and Cerf, Nicolas J. and Ralph, Timothy C. and Shapiro, Jeffrey H. and Lloyd, Seth},
  journal = {Rev. Mod. Phys.},
  volume = {84},
  issue = {2},
  pages = {621--669},
  numpages = {0},
  year = {2012},
  month = {May},
  publisher = {American Physical Society},
  doi = {10.1103/RevModPhys.84.621},
  url = {https://link.aps.org/doi/10.1103/RevModPhys.84.621}
}

@article{ku_generating_2015,
  title = {Generating and verifying entangled itinerant microwave fields with efficient and independent measurements},
  author = {Ku, H. S. and Kindel, W. F. and Mallet, F. and Glancy, S. and Irwin, K. D. and Hilton, G. C. and Vale, L. R. and Lehnert, K. W.},
  journal = {Phys. Rev. A},
  volume = {91},
  issue = {4},
  pages = {042305},
  numpages = {9},
  year = {2015},
  month = {Apr},
  publisher = {American Physical Society},
  doi = {10.1103/PhysRevA.91.042305},
  url = {https://link.aps.org/doi/10.1103/PhysRevA.91.042305}
}

@article{ockeloen_Noisless_2017,
  title = {Noiseless Quantum Measurement and Squeezing of Microwave Fields Utilizing Mechanical Vibrations},
  author = {Ockeloen-Korppi, C. F. and Damsk\"agg, E. and Pirkkalainen, J.-M. and Heikkil\"a, T. T. and Massel, F. and Sillanp\"a\"a, M. A.},
  journal = {Phys. Rev. Lett.},
  volume = {118},
  issue = {10},
  pages = {103601},
  numpages = {5},
  year = {2017},
  month = {Mar},
  publisher = {American Physical Society},
  doi = {10.1103/PhysRevLett.118.103601},
  url = {https://link.aps.org/doi/10.1103/PhysRevLett.118.103601}
}

@book{prabhakar2009spin,
    author = {Stancil, Daniel and Prabhakar, Anil},
    year = {2009},
    month = {01},
    pages = {},
    title = {Spin Waves: Theory and Applications},
    isbn = {978-0-387-77864-8},
    journal = {Spin Waves: Theory and Applications},
    doi = {10.1007/978-0-387-77865-5},
    publisher = {Springer},
    address = {New York, NY}
}

@article{eichler_observation_2011,
  title = {Observation of Two-Mode Squeezing in the Microwave Frequency Domain},
  author = {Eichler, C. and Bozyigit, D. and Lang, C. and Baur, M. and Steffen, L. and Fink, J. M. and Filipp, S. and Wallraff, A.},
  journal = {Phys. Rev. Lett.},
  volume = {107},
  issue = {11},
  pages = {113601},
  numpages = {5},
  year = {2011},
  month = {Sep},
  publisher = {American Physical Society},
  doi = {10.1103/PhysRevLett.107.113601},
  url = {https://link.aps.org/doi/10.1103/PhysRevLett.107.113601}
}

@article{morin_experimentally_2013,
  title = {Experimentally Accessing the Optimal Temporal Mode of Traveling Quantum Light States},
  author = {Morin, Olivier and Fabre, Claude and Laurat, Julien},
  journal = {Phys. Rev. Lett.},
  volume = {111},
  issue = {21},
  pages = {213602},
  numpages = {5},
  year = {2013},
  month = {Nov},
  publisher = {American Physical Society},
  doi = {10.1103/PhysRevLett.111.213602},
  url = {https://link.aps.org/doi/10.1103/PhysRevLett.111.213602}
}

@article{peugot_generating_2021,
  title = {Generating Two Continuous Entangled Microwave Beams Using a dc-Biased Josephson Junction},
  author = {Peugeot, A. and M\'enard, G. and Dambach, S. and Westig, M. and Kubala, B. and Mukharsky, Y. and Altimiras, C. and Joyez, P. and Vion, D. and Roche, P. and Esteve, D. and Milman, P. and Lepp\"akangas, J. and Johansson, G. and Hofheinz, M. and Ankerhold, J. and Portier, F.},
  journal = {Phys. Rev. X},
  volume = {11},
  issue = {3},
  pages = {031008},
  numpages = {9},
  year = {2021},
  month = {Jul},
  publisher = {American Physical Society},
  doi = {10.1103/PhysRevX.11.031008},
  url = {https://link.aps.org/doi/10.1103/PhysRevX.11.031008}
}

@article{anderson_instability_1955,
	title = {Instability in the {Motion} of {Ferromagnets} at {High} {Microwave} {Power} {Levels}},
	volume = {100},
	url = {https://link.aps.org/doi/10.1103/PhysRev.100.1788},
	doi = {10.1103/PhysRev.100.1788},
	number = {6},
	urldate = {2026-09-21},
	journal = {Phys. Rev.},
	publisher = {American Physical Society},
	author = {Anderson, P. W. and Suhl, H.},
	month = dec,
	year = {1955},
	pages = {1788--1789},
}

@book{lvov_wave_1994,
	address = {Berlin, Heidelberg},
	series = {Springer {Series} in {Nonlinear} {Dynamics}},
	title = {Wave {Turbulence} {Under} {Parametric} {Excitation}: {Applications} to {Magnets}},
	copyright = {https://www.springernature.com/gp/researchers/text-and-data-mining},
	isbn = {978-3-642-75297-1 978-3-642-75295-7},
	shorttitle = {Wave {Turbulence} {Under} {Parametric} {Excitation}},
	url = {https://link.springer.com/10.1007/978-3-642-75295-7},
	doi = {10.1007/978-3-642-75295-7},
	language = {english},
	urldate = {2026-09-21},
	publisher = {Springer},
	author = {L’vov, Victor S.},
	editor = {Calogero, F. and Fuchssteiner, B. and Rowlands, G. and Wadati, M. and Zakharov, V. E.},
	year = {1994},
}

@book{gurevich_magnetization_2020,
	address = {London},
	title = {Magnetization {Oscillations} and {Waves}},
	isbn = {978-0-13-874848-7},
	doi = {10.1201/9780138748487},
	publisher = {CRC Press},
	author = {Gurevich, A. G. and Melkov, G. A.},
	month = dec,
	year = {2020},
}

@article{boardman_three_1988,
  title = {Three- and four-magnon decay of nonlinear surface magnetostatic waves in thin ferromagnetic films},
  author = {Boardman, A. D. and Nikitov, S. A.},
  journal = {Phys. Rev. B},
  volume = {38},
  issue = {16},
  pages = {11444--11451},
  numpages = {0},
  year = {1988},
  month = {Dec},
  publisher = {American Physical Society},
  doi = {10.1103/PhysRevB.38.11444},
  url = {https://link.aps.org/doi/10.1103/PhysRevB.38.11444}
}

@article{ordonez-romero_thee_2009,
  title = {Three-magnon splitting and confluence processes for spin-wave excitations in yttrium iron garnet films: Wave vector selective Brillouin light scattering measurements and analysis},
  author = {Ord\'o\~nez-Romero, C\'esar L. and Kalinikos, Boris A. and Krivosik, Pavol and Tong, Wei and Kabos, Pavel and Patton, Carl E.},
  journal = {Phys. Rev. B},
  volume = {79},
  issue = {14},
  pages = {144428},
  numpages = {8},
  year = {2009},
  month = {Apr},
  publisher = {American Physical Society},
  doi = {10.1103/PhysRevB.79.144428},
  url = {https://link.aps.org/doi/10.1103/PhysRevB.79.144428}
}

@article{mohseni_controlling_2021,
  title = {Controlling the Nonlinear Relaxation of Quantized Propagating Magnons in Nanodevices},
  author = {Mohseni, M. and Wang, Q. and Heinz, B. and Kewenig, M. and Schneider, M. and Kohl, F. and L\"agel, B. and Dubs, C. and Chumak, A. V. and Pirro, P.},
  journal = {Phys. Rev. Lett.},
  volume = {126},
  issue = {9},
  pages = {097202},
  numpages = {6},
  year = {2021},
  month = {Mar},
  publisher = {American Physical Society},
  doi = {10.1103/PhysRevLett.126.097202},
  url = {https://link.aps.org/doi/10.1103/PhysRevLett.126.097202}
}

@article{krivosik_hamiltonian_2010,
  title = {Hamiltonian formulation of nonlinear spin-wave dynamics: Theory and applications},
  author = {Krivosik, Pavol and Patton, Carl E.},
  journal = {Phys. Rev. B},
  volume = {82},
  issue = {18},
  pages = {184428},
  numpages = {27},
  year = {2010},
  month = {Nov},
  publisher = {American Physical Society},
  doi = {10.1103/PhysRevB.82.184428},
  url = {https://link.aps.org/doi/10.1103/PhysRevB.82.184428}
}

@article{demidov_generation_2011,
  title = {Generation of the second harmonic by spin waves propagating in microscopic stripes},
  author = {Demidov, V. E. and Kostylev, M. P. and Rott, K. and Krzysteczko, P. and Reiss, G. and Demokritov, S. O.},
  journal = {Phys. Rev. B},
  volume = {83},
  issue = {5},
  pages = {054408},
  numpages = {5},
  year = {2011},
  month = {Feb},
  publisher = {American Physical Society},
  doi = {10.1103/PhysRevB.83.054408},
  url = {https://link.aps.org/doi/10.1103/PhysRevB.83.054408}
}

@article{chumak_magnon_2015,
	title = {Magnon spintronics},
	volume = {11},
	copyright = {2014 Springer Nature Limited},
	issn = {1745-2481},
	url = {https://www.nature.com/articles/nphys3347},
	doi = {10.1038/nphys3347},
	language = {english},
	number = {6},
	urldate = {2026-09-21},
	journal = {Nat. Phys.},
	publisher = {Nature Publishing Group},
	author = {Chumak, A. V. and Vasyuchka, V. I. and Serga, A. A. and Hillebrands, B.},
	month = jun,
	year = {2015},
	pages = {453--461},
}

@article{pirro_advances_2021,
	title = {Advances in coherent magnonics},
	volume = {6},
	copyright = {2021 Springer Nature Limited},
	issn = {2058-8437},
	url = {https://www.nature.com/articles/s41578-021-00332-w},
	doi = {10.1038/s41578-021-00332-w},
	language = {english},
	number = {12},
	urldate = {2026-09-21},
	journal = {Nat. Rev. Mater.},
	publisher = {Nature Publishing Group},
	author = {Pirro, Philipp and Vasyuchka, Vitaliy I. and Serga, Alexander A. and Hillebrands, Burkard},
	month = dec,
	year = {2021},
	pages = {1114--1135},
}

@article{eichler_quantum_2014,
  title = {Quantum-Limited Amplification and Entanglement in Coupled Nonlinear Resonators},
  author = {Eichler, C. and Salathe, Y. and Mlynek, J. and Schmidt, S. and Wallraff, A.},
  journal = {Phys. Rev. Lett.},
  volume = {113},
  issue = {11},
  pages = {110502},
  numpages = {5},
  year = {2014},
  month = {Sep},
  publisher = {American Physical Society},
  doi = {10.1103/PhysRevLett.113.110502},
  url = {https://link.aps.org/doi/10.1103/PhysRevLett.113.110502}
}

@misc{mycroftquantum_2026,
      title={Quantum State Preparation of Ferromagnetic Magnons by Parametric Driving}, 
      author={Monika E. Mycroft and Rostyslav O. Serha and Andrii V. Chumak and Carlos Gonzalez-Ballestero},
      year={2026},
      archivePrefix={arXiv},
      primaryClass={cond-mat.mes-hall},
      url={https://arxiv.org/abs/2601.12833}, 
}

@article{mills_quantum2006,
author = {Mills, D.L.},
year = {2006},
month = {11},
pages = {16-23},
title = {Quantum theory of spin waves in finite samples},
volume = {306},
journal = {J. Magn. Magn. Mater.},
doi = {10.1016/j.jmmm.2006.02.267}
}

@article{zakharov_spin-wave_1975,
	title = {Spin-wave turbulence beyond the parametric excitationthreshold},
	volume = {17},
	issn = {0038-5670},
	url = {https://iopscience.iop.org/article/10.1070/PU1975v017n06ABEH004404},
	doi = {10.1070/PU1975v017n06ABEH004404},
	language = {english},
	number = {6},
	urldate = {2026-09-24},
	journal = {Sov. Phys. Usp.},
	publisher = {IOP Publishing},
	author = {Zakharov, V. E. and L'vov, V. S. and Starobinets, S. S.},
	month = jun,
	year = {1975},
	pages = {896},
}

@article{mohseni_parametric_2020,
	title = {Parametric {Generation} of {Propagating} {Spin} {Waves} in {Ultrathin} {Yttrium} {Iron} {Garnet} {Waveguides}},
	volume = {14},
	issn = {1862-6270},
	url = {https://onlinelibrary.wiley.com/doi/abs/10.1002/pssr.202000011},
	doi = {10.1002/pssr.202000011},
	language = {english},
	number = {4},
	urldate = {2024-12-18},
	journal = {Phys. Status Solidi RRL},
	author = {Mohseni, Morteza and Kewenig, Martin and Verba, Roman and Wang, Qi and Schneider, Michael and Heinz, Björn and Kohl, Felix and Dubs, Carsten and Lägel, Bert and Serga, Alexander A. and Hillebrands, Burkard and Chumak, Andrii V. and Pirro, Philipp},
	year = {2020},
	pages = {2000011},
}

@article{bracher_mode_2011,
	title = {Mode selective parametric excitation of spin waves in a {Ni81Fe19} microstripe},
	volume = {99},
	issn = {0003-6951},
	url = {https://doi.org/10.1063/1.3651506},
	doi = {10.1063/1.3651506},
	number = {16},
	urldate = {2026-09-24},
	journal = {Appl. Phys. Lett.},
	author = {Brächer, T. and Pirro, P. and Obry, B. and Leven, B. and Serga, A. A. and Hillebrands, B.},
	month = oct,
	year = {2011},
	pages = {162501},
}

@article{melkov_parametric_1989,
    author    = {Melkov, G. A. and Sholom, S. V.},
    title     = {Parametric excitation of spin waves by a surface magnetostatic wave},
    journal   = {Sov. Phys. JETP},
    volume    = {69},
    number    = {2},
    pages     = {403--407},
    year      = {1989},
    url     = {https://jetp.ras.ru/cgi-bin/e/index/r/96/2/p712?a=list}
}

@article{bagada_parametric_1997,
  title = {Parametric Interaction of Dipolar Spin Wave Solitons with Localized Electromagnetic Pumping},
  author = {Bagada, A. V. and Melkov, G. A. and Serga, A. A. and Slavin, A. N.},
  journal = {Phys. Rev. Lett.},
  volume = {79},
  issue = {11},
  pages = {2137--2140},
  numpages = {0},
  year = {1997},
  month = {Sep},
  publisher = {American Physical Society},
  doi = {10.1103/PhysRevLett.79.2137},
  url = {https://link.aps.org/doi/10.1103/PhysRevLett.79.2137}
}

@book{walls_quantum_2008,
	address = {Berlin, Heidelberg},
	title = {Quantum {Optics}},
	copyright = {http://www.springer.com/tdm},
	isbn = {978-3-540-28573-1},
	url = {http://link.springer.com/10.1007/978-3-540-28574-8},
	doi = {10.1007/978-3-540-28574-8},
	language = {english},
	urldate = {2026-09-24},
	publisher = {Springer},
	editor = {Walls, D.F. and Milburn, Gerard J.},
	year = {2008}
}

@article{duan_inseparability_2000,
  title = {Inseparability Criterion for Continuous Variable Systems},
  author = {Duan, Lu-Ming and Giedke, G. and Cirac, J. I. and Zoller, P.},
  journal = {Phys. Rev. Lett.},
  volume = {84},
  issue = {12},
  pages = {2722--2725},
  numpages = {0},
  year = {2000},
  month = {Mar},
  publisher = {American Physical Society},
  doi = {10.1103/PhysRevLett.84.2722},
  url = {https://link.aps.org/doi/10.1103/PhysRevLett.84.2722}
}

@article{simon_peres-horodecki_2000,
  title = {Peres-Horodecki Separability Criterion for Continuous Variable Systems},
  author = {Simon, R.},
  journal = {Phys. Rev. Lett.},
  volume = {84},
  issue = {12},
  pages = {2726--2729},
  numpages = {0},
  year = {2000},
  month = {Mar},
  publisher = {American Physical Society},
  doi = {10.1103/PhysRevLett.84.2726},
  url = {https://link.aps.org/doi/10.1103/PhysRevLett.84.2726}
}

@article{Qin_propagating_2018,
  title = {Propagating spin waves in nanometer-thick yttrium iron garnet films: Dependence on wave vector, magnetic field strength, and angle},
  author = {Qin, Huajun and H\"am\"al\"ainen, Sampo J. and Arjas, Kristian and Witteveen, Jorn and van Dijken, Sebastiaan},
  journal = {Phys. Rev. B},
  volume = {98},
  issue = {22},
  pages = {224422},
  numpages = {8},
  year = {2018},
  month = {Dec},
  publisher = {American Physical Society},
  doi = {10.1103/PhysRevB.98.224422},
  url = {https://link.aps.org/doi/10.1103/PhysRevB.98.224422}
}

@article{jiang_thin-film_2026,
	title = {Thin-film magnomechanics in the low gigahertz regime},
	volume = {139},
	issn = {0021-8979},
	url = {https://doi.org/10.1063/5.0315911},
	doi = {10.1063/5.0315911},
	number = {11},
	urldate = {2026-09-24},
	journal = {J. Appl. Phys.},
	author = {Jiang, Yu and Yan, Zixin and Huang, Yizhong and Zhang, Xufeng},
	month = mar,
	year = {2026},
	pages = {113904},
}

@book{gardiner_quantum_2004,
  title     = {Quantum Noise: A Handbook of Markovian and Non-Markovian Quantum Stochastic Methods with Applications to Quantum Optics},
  author    = {Gardiner, Crispin W. and Zoller, Peter},
  edition   = {3rd},
  series    = {Springer Series in Synergetics},
  year      = {2004},
  publisher = {Springer Science \& Business Media},
  address   = {Berlin, Heidelberg},
  isbn      = {978-3-540-22301-6},
  doi       = {10.1007/978-3-662-10803-1}
}

@ARTICLE{chumak_roadmap_2022,
  author={Chumak, A. V. and Kabos, P. and Wu, M. and Abert, C. and Adelmann, C. and Adeyeye, A. O. and Åkerman, J. and Aliev, F. G. and Anane, A. and Awad, A. and Back, C. H. and Barman, A. and Bauer, G. E. W. and Becherer, M. and Beginin, E. N. and Bittencourt, V. A. S. V. and Blanter, Y. M. and Bortolotti, P. and Boventer, I. and Bozhko, D. A. and Bunyaev, S. A. and Carmiggelt, J. J. and Cheenikundil, R. R. and Ciubotaru, F. and Cotofana, S. and Csaba, G. and Dobrovolskiy, O. V. and Dubs, C. and Elyasi, M. and Fripp, K. G. and Fulara, H. and Golovchanskiy, I. A. and Gonzalez-Ballestero, C. and Graczyk, P. and Grundler, D. and Gruszecki, P. and Gubbiotti, G. and Guslienko, K. and Haldar, A. and Hamdioui, S. and Hertel, R. and Hillebrands, B. and Hioki, T. and Houshang, A. and Hu, C.-M. and Huebl, H. and Huth, M. and Iacocca, E. and Jungfleisch, M. B. and Kakazei, G. N. and Khitun, A. and Khymyn, R. and Kikkawa, T. and Kläui, M. and Klein, O. and Kłos, J. W. and Knauer, S. and Koraltan, S. and Kostylev, M. and Krawczyk, M. and Krivorotov, I. N. and Kruglyak, V. V. and Lachance-Quirion, D. and Ladak, S. and Lebrun, R. and Li, Y. and Lindner, M. and Macêdo, R. and Mayr, S. and Melkov, G. A. and Mieszczak, S. and Nakamura, Y. and Nembach, H. T. and Nikitin, A. A. and Nikitov, S. A. and Novosad, V. and Otálora, J. A. and Otani, Y. and Papp, A. and Pigeau, B. and Pirro, P. and Porod, W. and Porrati, F. and Qin, H. and Rana, B. and Reimann, T. and Riente, F. and Romero-Isart, O. and Ross, A. and Sadovnikov, A. V. and Safin, A. R. and Saitoh, E. and Schmidt, G. and Schultheiss, H. and Schultheiss, K. and Serga, A. A. and Sharma, S. and Shaw, J. M. and Suess, D. and Surzhenko, O. and Szulc, K. and Taniguchi, T. and Urbánek, M. and Usami, K. and Ustinov, A. B. and van der Sar, T. and van Dijken, S. and Vasyuchka, V. I. and Verba, R. and Kusminskiy, S. Viola and Wang, Q. and Weides, M. and Weiler, M. and Wintz, S. and Wolski, S. P. and Zhang, X.},
  journal={IEEE Transactions on Magnetics}, 
  title={Advances in Magnetics Roadmap on Spin-Wave Computing}, 
  year={2022},
  volume={58},
  number={6},
  pages={1-72},
  doi={10.1109/TMAG.2022.3149664}
  }

@book{barnett_methods_1997,
  author    = {Barnett, Stephen M. and Radmore, Paul M.},
  title     = {Methods in Theoretical Quantum Optics},
  publisher = {Oxford University Press},
  address   = {Oxford},
  year      = {1997},
  series    = {Oxford Series in Optical and Imaging Sciences},
  volume    = {15},
  isbn      = {978-0-19-856361-7}
}

@article{collett_input_1984,
  author    = {Collett, M. J. and Gardiner, C. W.},
  title     = {Squeezing of intracavity and traveling-wave light fields 
               produced in parametric amplification},
  journal   = {Phys. Rev. A},
  volume    = {30},
  pages     = {1386},
  year      = {1984},
  doi       = {10.1103/PhysRevA.30.1386}
}

@article{ciubotaru_all_2016,
  author    = {Ciubotaru, F. and Devolder, T. and Manfrini, M. 
               and Adelmann, C. and Radu, I.},
  title     = {All electrical propagating spin wave spectroscopy 
               with broadband wavevector capability},
  journal   = {Appl. Phys. Lett.},
  volume    = {109},
  pages     = {012403},
  year      = {2016},
  doi       = {10.1063/1.4955030}
}

@book{landau1980statisticalII,
    author       = {L. D. Landau and E. M. Lifshitz and L. P. Pitaevskii},
    title        = {Statistical Physics, Part 2},
    subtitle     = {Theory of the Condensed State},
    series       = {Course of Theoretical Physics},
    volume       = {9},
    publisher    = {Pergamon Press},
    address      = {Oxford, UK},
    year         = {1980},
    isbn         = {9780080230733}
}

@article{fedorov_finite-time_2018,
	title = {Finite-time quantum entanglement in propagating squeezed microwaves},
	volume = {8},
	issn = {2045-2322},
	url = {https://doi.org/10.1038/s41598-018-24742-z},
	doi = {10.1038/s41598-018-24742-z},
	number = {1},
	journal = {Sci. Rep.},
	author = {Fedorov, K. G. and Pogorzalek, S. and Las Heras, U. and Sanz, M. and Yard, P. and Eder, P. and Fischer, M. and Goetz, J. and Xie, E. and Inomata, K. and Nakamura, Y. and Di Candia, R. and Solano, E. and Marx, A. and Deppe, F. and Gross, R.},
	year = {2018},
	pages = {6416},
}

@article{Leppakangas_nonclassical_2013,
  title = {Nonclassical Photon Pair Production in a Voltage-Biased Josephson Junction},
  author = {Lepp{\"a}kangas, Juha and Johansson, G{\"o}ran and Marthaler, Michael and Fogelstr{\"o}m, Mikael},
  journal = {Phys. Rev. Lett.},
  volume = {110},
  issue = {26},
  pages = {267004},
  numpages = {5},
  year = {2013},
  month = {Jun},
  publisher = {American Physical Society},
  doi = {10.1103/PhysRevLett.110.267004},
  url = {https://link.aps.org/doi/10.1103/PhysRevLett.110.267004}
}

@article{barzanjeh_stationary_2019,
	title = {Stationary entangled radiation from micromechanical motion},
	volume = {570},
	copyright = {2019 The Author(s), under exclusive licence to Springer Nature Limited},
	issn = {1476-4687},
	url = {https://www.nature.com/articles/s41586-019-1320-2},
	doi = {10.1038/s41586-019-1320-2},
	language = {english},
	number = {7762},
	urldate = {2026-11-11},
	journal = {Nature},
	publisher = {Nature Publishing Group},
	author = {Barzanjeh, S. and Redchenko, E. S. and Peruzzo, M. and Wulf, M. and Lewis, D. P. and Arnold, G. and Fink, J. M.},
	month = jun,
	year = {2019},
	pages = {480--483}
}

@misc{mohamed_path-entangled_2024,
	title = {Path-entangled radiation from kinetic inductance amplifier},
	url = {http://arxiv.org/abs/2406.13239},
	doi = {10.48550/arXiv.2406.13239},
	urldate = {2026-11-11},
	publisher = {arXiv},
	author = {Mohamed, Abdul and Barzanjeh, Shabir},
	month = jun,
	year = {2024},
	note = {arXiv:2406.13239 [quant-ph]},
}

@article{Xu_Magnetic_2023,
  title = {Magnetic Field-Resilient Quantum-Limited Parametric Amplifier},
  author = {Xu, Mingrui and Cheng, Risheng and Wu, Yufeng and Liu, Gangqiang and Tang, Hong X.},
  journal = {PRX Quantum},
  volume = {4},
  issue = {1},
  pages = {010322},
  numpages = {9},
  year = {2023},
  month = {Feb},
  publisher = {American Physical Society},
  doi = {10.1103/PRXQuantum.4.010322},
  url = {https://link.aps.org/doi/10.1103/PRXQuantum.4.010322}
}

@article{Wu_junction-free_2024,
  title = {Junction-free microwave two-mode radiation from a kinetic inductance nanowire},
  author = {Wu, Yufeng and Xu, Mingrui and Tang, Hong X.},
  journal = {Phys. Rev. Appl.},
  volume = {21},
  issue = {1},
  pages = {014029},
  numpages = {9},
  year = {2024},
  month = {Jan},
  publisher = {American Physical Society},
  doi = {10.1103/PhysRevApplied.21.014029},
  url = {https://link.aps.org/doi/10.1103/PhysRevApplied.21.014029}
}

@article{Forgues_experimental_2015,
  title = {Experimental Violation of Bell-like Inequalities By Electronic Shot Noise},
  author = {Forgues, Jean-Charles and Lupien, Christian and Reulet, Bertrand},
  journal = {Phys. Rev. Lett.},
  volume = {114},
  issue = {13},
  pages = {130403},
  numpages = {5},
  year = {2015},
  month = {Apr},
  publisher = {American Physical Society},
  doi = {10.1103/PhysRevLett.114.130403},
  url = {https://link.aps.org/doi/10.1103/PhysRevLett.114.130403}
}

@article{Di_high-efficiency_2023,
author = {Ke Di and Shuai Tan and Liyong Wang and Anyu Cheng and Xi Wang and Yu Liu and Jiajia Du},
journal = {Opt. Express},
number = {18},
pages = {29491--29503},
publisher = {Optica Publishing Group},
title = {High-efficiency entanglement of microwave fields in cavity opto-magnomechanical systems},
volume = {31},
month = {Aug},
year = {2023},
url = {https://opg.optica.org/oe/abstract.cfm?URI=oe-31-18-29491},
doi = {10.1364/OE.495656},
}

@article{mahmoud_introduction_2020,
	title = {Introduction to spin wave computing},
	volume = {128},
	issn = {0021-8979},
	url = {https://doi.org/10.1063/5.0019328},
	doi = {10.1063/5.0019328},
	number = {16},
	urldate = {2026-11-11},
	journal = {J. Appl. Phys.},
	author = {Mahmoud, Abdulqader and Ciubotaru, Florin and Vanderveken, Frederic and Chumak, Andrii V. and Hamdioui, Said and Adelmann, Christoph and Cotofana, Sorin},
	month = oct,
	year = {2020},
	pages = {161101},
}

@article{demokritov_boseeinstein_2006,
	title = {Bose–{Einstein} condensation of quasi-equilibrium magnons at room temperature under pumping},
	volume = {443},
	copyright = {2006 Springer Nature Limited},
	issn = {1476-4687},
	url = {https://www.nature.com/articles/nature05117},
	doi = {10.1038/nature05117},
	language = {english},
	number = {7110},
	urldate = {2026-11-11},
	journal = {Nature},
	publisher = {Nature Publishing Group},
	author = {Demokritov, S. O. and Demidov, V. E. and Dzyapko, O. and Melkov, G. A. and Serga, A. A. and Hillebrands, B. and Slavin, A. N.},
	month = sep,
	year = {2006},
	pages = {430--433},
}

@article{Serpico_quantum_2025,
  title = {Quantum micromagnetic theory of magnons in finite nanostructures},
  author = {Serpico, Claudio and Perna, Salvatore and d'Aquino, Massimiliano},
  journal = {Phys. Rev. B},
  volume = {111},
  issue = {18},
  pages = {184410},
  numpages = {17},
  year = {2025},
  month = {May},
  publisher = {American Physical Society},
  doi = {10.1103/PhysRevB.111.184410},
  url = {https://link.aps.org/doi/10.1103/PhysRevB.111.184410}
}

@article{Kubo_generalized_1962,
    title = {Generalized Cumulant Expansion Method},
    volume = {17},
    ISSN = {1347-4073},
    DOI = {10.1143/jpsj.17.1100},
    number = {7},
    journal = {J. Phys. Soc. Jpn.},
    publisher = {Physical Society of Japan},
    author = {Kubo,  Ryogo},
    year = {1962},
    month = {Jul},
    pages = {1100--1120}
}

@article{Kadanoff_meanfield_2009,
	title = {More is the {Same}; {Phase} {Transitions} and {Mean} {Field} {Theories}},
	volume = {137},
	issn = {1572-9613},
	url = {https://doi.org/10.1007/s10955-009-9814-1},
	doi = {10.1007/s10955-009-9814-1},
	number = {5},
	journal = {J. Stat. Phys.},
	author = {Kadanoff, Leo P.},
	month = {Dec},
	year = {2009},
	pages = {777--797},
}

@article{Parker_degenerate_2022,
  title = {Degenerate Parametric Amplification via Three-Wave Mixing Using Kinetic Inductance},
  author = {Parker, Daniel J. and Savytskyi, Mykhailo and Vine, Wyatt and Laucht, Arne and Duty, Timothy and Morello, Andrea and Grimsmo, Arne L. and Pla, Jarryd J.},
  journal = {Phys. Rev. Appl.},
  volume = {17},
  issue = {3},
  pages = {034064},
  numpages = {35},
  year = {2022},
  month = {Mar},
  publisher = {American Physical Society},
  doi = {10.1103/PhysRevApplied.17.034064},
  url = {https://link.aps.org/doi/10.1103/PhysRevApplied.17.034064}
}

@article{chumak_magnon_2014,
	title = {Magnon transistor for all-magnon data processing},
	volume = {5},
	copyright = {2014 The Author(s)},
	issn = {2041-1723},
	url = {https://www.nature.com/articles/ncomms5700},
	doi = {10.1038/ncomms5700},
	language = {english},
	number = {1},
	urldate = {2026-11-12},
	journal = {Nat. Commun.},
	publisher = {Nature Publishing Group},
	author = {Chumak, Andrii V. and Serga, Alexander A. and Hillebrands, Burkard},
	month = aug,
	year = {2014},
	pages = {4700},
}

@article{wang_magnonic_2020,
	title = {A magnonic directional coupler for integrated magnonic half-adders},
	volume = {3},
	copyright = {2020 The Author(s), under exclusive licence to Springer Nature Limited},
	issn = {2520-1131},
	url = {https://www.nature.com/articles/s41928-020-00485-6},
	doi = {10.1038/s41928-020-00485-6},
	language = {english},
	number = {12},
	urldate = {2026-11-12},
	journal = {Nat. Electron.},
	publisher = {Nature Publishing Group},
	author = {Wang, Q. and Kewenig, M. and Schneider, M. and Verba, R. and Kohl, F. and Heinz, B. and Geilen, M. and Mohseni, M. and Lägel, B. and Ciubotaru, F. and Adelmann, C. and Dubs, C. and Cotofana, S. D. and Dobrovolskiy, O. V. and Brächer, T. and Pirro, P. and Chumak, A. V.},
	month = dec,
	year = {2020},
	pages = {765--774},
}

@misc{bessonov_frequency-domain_2026,
	title = {Frequency-domain spin-wave logic gate based on three-magnon confluence},
	url = {http://arxiv.org/abs/2611.02946},
	doi = {10.48550/arXiv.2611.02946},
	urldate = {2026-11-12},
	publisher = {arXiv},
	author = {Bessonov, Vladimir D. and Lake, Stephanie R. and Mohapatra, Bikash Das and Bensmann, Jannis and Schmidt, Georg and Bratschitsch, Rudolf and Demokritov, Sergej O. and Demidov, Vladislav E.},
	month = nov,
	year = {2026},
	note = {arXiv:2611.02946 [cond-mat.mes-hall]},
}

@article{zenbaa_realization_2025,
	title = {Realization of inverse-design magnonic logic gates},
	volume = {11},
	url = {https://www.science.org/doi/10.1126/sciadv.adu9032},
	doi = {10.1126/sciadv.adu9032},
	number = {21},
	urldate = {2026-11-12},
	journal = {Sci. Adv.},
	publisher = {American Association for the Advancement of Science},
	author = {Zenbaa, Noura and Majcen, Fabian and Abert, Claas and Bruckner, Florian and Mauser, Norbert J. and Schrefl, Thomas and Wang, Qi and Suess, Dieter and Chumak, Andrii V.},
	month = may,
	year = {2025},
	pages = {eadu9032},
}

@article{papp_nanoscale_2021,
	title = {Nanoscale neural network using non-linear spin-wave interference},
	volume = {12},
	copyright = {2021 The Author(s)},
	issn = {2041-1723},
	url = {https://www.nature.com/articles/s41467-021-26711-z},
	doi = {10.1038/s41467-021-26711-z},
	language = {english},
	number = {1},
	urldate = {2026-11-12},
	journal = {Nat. Commun.},
	publisher = {Nature Publishing Group},
	author = {Papp, \'Adám and Porod, Wolfgang and Csaba, Gyorgy},
	month = nov,
	year = {2021},
	pages = {6422},
}

@misc{guo_integrated_2026,
	title = {Integrated magnonic neural circuits based on nonlinear wave neurons},
	url = {http://arxiv.org/abs/2606.11703},
	doi = {10.48550/arXiv.2606.11703},
	urldate = {2026-11-12},
	publisher = {arXiv},
	author = {Guo, Mengying and Jing, Xudong and Davidkova, Kristýna and Verba, Roman and Zhou, Zhenyu and Guo, Xueyu and Dubs, Carsten and Gao, Chuan and Rao, Yiheng and Cai, Kaiming and Li, Jing and Pirro, Philipp and Chumak, Andrii V. and Wang, Qi},
	month = jun,
	year = {2026},
	note = {arXiv:2606.11703 [cond-mat.mtrl-sci]},
}

@misc{breitbach_all-magnonic_2026,
	title = {An all-magnonic neuron with tunable fading memory},
	url = {http://arxiv.org/abs/2509.18321},
	doi = {10.48550/arXiv.2509.18321},
	urldate = {2026-11-12},
	publisher = {arXiv},
	author = {Breitbach, David and Bechberger, Moritz and Mortada, Hanadi and Heinz, Björn and Verba, Roman and Wang, Qi and Dubs, Carsten and Carpentieri, Mario and Finocchio, Giovanni and Rodrigues, Davi and Hamadeh, Alexandre Abbass and Pirro, Philipp},
	month = mar,
	year = {2026},
	note = {arXiv:2509.18321 [cond-mat.mtrl-sci]},
}

@article{Watt_Implementing_2021,
  title = {Implementing a Magnonic Reservoir Computer Model Based on Time-Delay Multiplexing},
  author = {Watt, Stuart and Kostylev, Mikhail and Ustinov, Alexey B. and Kalinikos, Boris A.},
  journal = {Phys. Rev. Appl.},
  volume = {15},
  issue = {6},
  pages = {064060},
  numpages = {14},
  year = {2021},
  month = {Jun},
  publisher = {American Physical Society},
  doi = {10.1103/PhysRevApplied.15.064060},
  url = {https://link.aps.org/doi/10.1103/PhysRevApplied.15.064060}
}

@article{korber_pattern_2023,
	title = {Pattern recognition in reciprocal space with a magnon-scattering reservoir},
	volume = {14},
	copyright = {2023 The Author(s)},
	issn = {2041-1723},
	url = {https://www.nature.com/articles/s41467-023-39452-y},
	doi = {10.1038/s41467-023-39452-y},
	language = {english},
	number = {1},
	urldate = {2026-11-12},
	journal = {Nat. Commun.},
	publisher = {Nature Publishing Group},
	author = {K{\"o}rber, Lukas and Heins, Christopher and Hula, Tobias and Kim, Joo-Von and Thlang, Sonia and Schultheiss, Helmut and Fassbender, Jürgen and Schultheiss, Katrin},
	month = jul,
	year = {2023},
	pages = {3954},
}

@article{Xiong_predicting_2026,
  title = {Predicting the future with magnons: Forecasting chaotic time series with reservoir computing},
  author = {Xiong, Zeling and Heins, Christopher and Devolder, Thibaut and Kammerbauer, Fabian and Kl{\"a}ui, Mathias and Fassbender, J{\"u}rgen and Schultheiss, Helmut and Schultheiss, Katrin},
  journal = {Phys. Rev. Appl.},
  volume = {25},
  issue = {4},
  pages = {044047},
  numpages = {12},
  year = {2026},
  month = {Apr},
  publisher = {American Physical Society},
  doi = {10.1103/ffz7-p6vc},
  url = {https://link.aps.org/doi/10.1103/ffz7-p6vc}
}

@article{hofinger_k-selective_2026,
	title = {k-{Selective} {Electrical}-to-{Magnon} {Transduction} with {Finite}-{Element}-{Resolved} {Sub}-{Micron} {Nanoantennas}},
	volume = {5},
	issn = {2751-1200},
	url = {https://onlinelibrary.wiley.com/doi/abs/10.1002/apxr.202500211},
	doi = {10.1002/apxr.202500211},
	language = {english},
	number = {6},
	urldate = {2026-11-12},
	journal = {Adv. Phys. Res.},
	author = {Höfinger, Andreas and Voronov, Andrey A. and Schmoll, David and Koraltan, Sabri and Bruckner, Florian and Abert, Claas and Suess, Dieter and Lindner, Morris and Reimann, Timmy and Dubs, Carsten and Chumak, Andrii V. and Knauer, Sebastian},
	year = {2026},
	pages = {e00211},
}

@article{casulleras_generation_2023,
  title = {Generation of Spin-Wave Pulses by Inverse Design},
  author = {Casulleras, S. and Knauer, S. and Wang, Q. and Romero-Isart, O. and Chumak, A.V. and Gonzalez-Ballestero, C.},
  journal = {Phys. Rev. Appl.},
  volume = {19},
  issue = {6},
  pages = {064085},
  numpages = {10},
  year = {2023},
  month = {Jun},
  publisher = {American Physical Society},
  doi = {10.1103/PhysRevApplied.19.064085},
  url = {https://link.aps.org/doi/10.1103/PhysRevApplied.19.064085}
}

@article{li_strong_2026,
  title = {Strong Coupling between Propagating Spin Waves and Microwave Photons in a Superconducting Resonator},
  author = {Li, Yi and Lim, Jinho and Wang, Xingzhi and Polakovic, Tomas and Kiehl, Carissa and Song, Moojune and Van, Phuoc Cao and Divan, Ralu and Welp, Ulrich and Phatak, Charudatta and Jeong, Jong-Ryul and Kim, Kab-Jin and Zuo, Jian-Min and Hoffmann, Axel and Novosad, Valentine},
  journal = {Phys. Rev. Lett.},
  volume = {137},
  issue = {11},
  pages = {116706},
  numpages = {8},
  year = {2026},
  month = {Sep},
  publisher = {American Physical Society},
  doi = {10.1103/8t91-w9nn},
  url = {https://link.aps.org/doi/10.1103/8t91-w9nn}
}

@article{jiang_integrating_2023,
	title = {Integrating magnons for quantum information},
	volume = {123},
	issn = {0003-6951},
	url = {https://doi.org/10.1063/5.0157520},
	doi = {10.1063/5.0157520},
	number = {13},
	urldate = {2026-11-12},
	journal = {Appl. Phys. Lett.},
	author = {Jiang, Zhihao and Lim, Jinho and Li, Yi and Pfaff, Wolfgang and Lo, Tzu-Hsiang and Qian, Jiangchao and Schleife, André and Zuo, Jian-Min and Novosad, Valentine and Hoffmann, Axel},
	month = sep,
	year = {2023},
	pages = {130501},
}

@article{li_hybrid_2020,
	title = {Hybrid magnonics: {Physics}, circuits, and applications for coherent information processing},
	volume = {128},
	issn = {0021-8979},
	shorttitle = {Hybrid magnonics},
	url = {https://doi.org/10.1063/5.0020277},
	doi = {10.1063/5.0020277},
	number = {13},
	urldate = {2026-11-12},
	journal = {J. Appl. Phys.},
	author = {Li, Yi and Zhang, Wei and Tyberkevych, Vasyl and Kwok, Wai-Kwong and Hoffmann, Axel and Novosad, Valentine},
	month = oct,
	year = {2020},
	pages = {130902},
}

@article{serha_ultralong-living_2026,
	title = {Ultralong-living magnons in the quantum limit},
	volume = {12},
	url = {https://www.science.org/doi/full/10.1126/sciadv.aee2344},
	doi = {10.1126/sciadv.aee2344},
	number = {18},
	urldate = {2026-11-16},
	journal = {Sci. Adv.},
	publisher = {American Association for the Advancement of Science},
	author = {Serha, Rostyslav O. and McAllister, Kaitlin H. and Majcen, Fabian and Knauer, Sebastian and Reimann, Timmy and Dubs, Carsten and Melkov, Gennadii A. and Serga, Alexander A. and Tyberkevych, Vasyl S. and Chumak, Andrii V. and Bozhko, Dmytro A.},
	month = may,
	year = {2026},
	pages = {eaee2344},
}

@misc{schmoll_brillouin_2026,
	title = {Brillouin {Light} {Scattering} {Spectroscopy} of {Propagating} {Magnons} at {Sub}-{Kelvin} {Temperatures}},
	url = {http://arxiv.org/abs/2606.25834},
	doi = {10.48550/arXiv.2606.25834},
	urldate = {2026-11-12},
	publisher = {arXiv},
	author = {Schmoll, David and Kuznetsov, Nikolai and Rehberger, Phillip and Vilsmeier, Franz and Verba, Roman and Slobodianiuk, Denys and Serha, Rostyslav O. and Levchenko, Khrystyna O. and Dijken, Sebastiaan van and Chumak, Andrii V. and Knauer, Sebastian},
	month = jun,
	year = {2026},
	note = {arXiv:2606.25834 [cond-mat.other]},
}

@article{serha_magnetic_2024,
	title = {Magnetic anisotropy and {GGG} substrate stray field in {YIG} films down to millikelvin temperatures},
	volume = {2},
	copyright = {2024 The Author(s)},
	issn = {2948-2119},
	url = {https://www.nature.com/articles/s44306-024-00030-7},
	doi = {10.1038/s44306-024-00030-7},
	language = {english},
	number = {1},
	urldate = {2026-11-12},
	journal = {npj Spintronics},
	publisher = {Nature Publishing Group},
	author = {Serha, Rostyslav O. and Voronov, Andrey A. and Schmoll, David and Verba, Roman and Levchenko, Khrystyna O. and Koraltan, Sabri and Davídková, Kristýna and Budinská, Barbora and Wang, Qi and Dobrovolskiy, Oleksandr V. and Urbánek, Michal and Lindner, Morris and Reimann, Timmy and Dubs, Carsten and Gonzalez-Ballestero, Carlos and Abert, Claas and Suess, Dieter and Bozhko, Dmytro A. and Knauer, Sebastian and Chumak, Andrii V.},
	month = jul,
	year = {2024},
	pages = {29},
}

@unpublished{Meese_unpublished,
    author = {Meese, B. T. and Gonzalez-Ballestero, C},
    note = {[Manuscript in preparation]},
    year = {2026}
}

@article{kani_squeezed_2025,
	title = {Squeezed microwave and magnonic frequency combs},
	volume = {2},
	issn = {2835-0103},
	url = {https://doi.org/10.1063/5.0245193},
	doi = {10.1063/5.0245193},
	number = {1},
	urldate = {2026-11-12},
	journal = {APL Quantum},
	author = {Kani, A. and Hatifi, M. and Twamley, J.},
	month = jan,
	year = {2025},
	pages = {016112},
}

@article{safonov_nonlinear_1996,
	title = {Nonlinear radiation damping of parametrically excited spin waves},
	volume = {161},
	issn = {0304-8853},
	url = {https://www.sciencedirect.com/science/article/pii/S0304885396012851},
	doi = {https://doi.org/10.1016/S0304-8853(96)01285-1},
	journal = {J. Magn. Magn. Mater.},
	author = {Safonov, V. L. and Yamazaki, H.},
	year = {1996},
	pages = {275--281},
}

@article{liang-sheng_photonmagnon_2005,
	title = {Photon–magnon interaction process: a mechanism of resonance linewidths of ferromagnet},
	volume = {14},
	url = {https://doi.org/10.1088/1009-1963/14/1/020},
	doi = {10.1088/1009-1963/14/1/020},
	number = {1},
	journal = {Chin. Phys.},
	author = {Liang-Sheng, Li and Qing-Fan, Shi},
	month = jan,
	year = {2005},
	pages = {110},
}

@article{qing-fan_nonlinear_2003,
	title = {Nonlinear interaction of resonator photons with parametric magnon pairs},
	volume = {12},
	url = {https://doi.org/10.1088/1009-1963/12/4/309},
	doi = {10.1088/1009-1963/12/4/309},
	number = {4},
	journal = {Chin. Phys.},
	author = {Qing-Fan, Shi and Xue-Qun, Yan and Xue-Guan, Liu},
	month = apr,
	year = {2003},
	pages = {394},
}

@article{jander_parametric_2025,
	title = {Parametric pumping of coherent spin waves by surface acoustic waves},
	volume = {137},
	issn = {0021-8979},
	url = {https://doi.org/10.1063/5.0251252},
	doi = {10.1063/5.0251252},
	number = {7},
	urldate = {2026-11-12},
	journal = {J. Appl. Phys.},
	author = {Jander, Albrecht and Dhagat, Pallavi and Rivard, Carson},
	month = feb,
	year = {2025},
	pages = {073902},
}

@article{rivard_sustained_2025,
	title = {Sustained amplification of coherent spin waves by parametric pumping with surface acoustic waves},
	volume = {3},
	copyright = {2025 The Author(s)},
	issn = {2948-2119},
	url = {https://www.nature.com/articles/s44306-025-00115-x},
	doi = {10.1038/s44306-025-00115-x},
	language = {english},
	number = {1},
	urldate = {2026-11-12},
	journal = {npj Spintronics},
	publisher = {Nature Publishing Group},
	author = {Rivard, Carson and Jander, Albrecht and Dhagat, Pallavi},
	month = dec,
	year = {2025},
	pages = {49},
}

@article{rivard_parametric_2025,
	title = {Parametric instability of forward volume spin waves pumped by surface acoustic waves in yttrium iron garnet films},
	volume = {127},
	issn = {0003-6951},
	url = {https://doi.org/10.1063/5.0280832},
	doi = {10.1063/5.0280832},
	number = {4},
	urldate = {2026-11-12},
	journal = {Appl. Phys. Lett.},
	author = {Rivard, C. and Jander, A. and Dhagat, P.},
	month = jul,
	year = {2025},
	pages = {042408},
}

@article{Chowdhury_nondegenerate_2017,
  author={Chowdhury, Pratim and Jander, Albrecht and Dhagat, Pallavi},
  journal={IEEE Magn. Lett.}, 
  title={Nondegenerate Parametric Pumping of Spin Waves by Acoustic Waves}, 
  year={2017},
  volume={8},
  number={},
  pages={1-4},
  doi={10.1109/LMAG.2017.2737962}}

@article{zhang_cavity_2016,
	title = {Cavity magnomechanics},
	volume = {2},
	url = {https://www.science.org/doi/10.1126/sciadv.1501286},
	doi = {10.1126/sciadv.1501286},
	number = {3},
	urldate = {2023-08-11},
	journal = {Sci. Adv.},
	publisher = {American Association for the Advancement of Science},
	author = {Zhang, Xufeng and Zou, Chang-Ling and Jiang, Liang and Tang, Hong X.},
	month = mar,
	year = {2016},
	pages = {e1501286},
}

@article{Bittencourt_Magnomechanical_2023,
  title = {Magnomechanical backaction corrections due to coupling to higher-order Walker modes and Kerr nonlinearities},
  author = {Bittencourt, V. A. S. V. and Potts, C. A. and Huang, Y. and Davis, J. P. and Viola Kusminskiy, S.},
  journal = {Phys. Rev. B},
  volume = {107},
  issue = {14},
  pages = {144411},
  numpages = {18},
  year = {2023},
  month = {Apr},
  publisher = {American Physical Society},
  doi = {10.1103/PhysRevB.107.144411},
  url = {https://link.aps.org/doi/10.1103/PhysRevB.107.144411}
}

@article{Cherkasskii_Theory_2025,
    author = {Cherkasskii, M. and Engelhardt, F. and Müller, M. and Weber, J. and Althammer, M. and Goennenwein, S. T. B. and Huebl, H. and Viola Kusminskiy, S.},
    title = {Theory of polarization-dependent phonon pumping in ferromagnetic/non-magnetic bilayers},
    journal = {J. Appl. Phys.},
    volume = {138},
    number = {4},
    pages = {043906},
    year = {2025},
    month = {07},
    issn = {0021-8979},
    doi = {10.1063/5.0273075},
    url = {https://doi.org/10.1063/5.0273075},
}

@article{geilen_parametric_2025,
	title = {Parametric {Excitation} and {Instabilities} of {Spin} {Waves} {Driven} by {Surface} {Acoustic} {Waves}},
	volume = {4},
	issn = {2751-1200},
	url = {https://advanced.onlinelibrary.wiley.com/doi/10.1002/apxr.202400086},
	doi = {10.1002/apxr.202400086},
	language = {english},
	number = {1},
	urldate = {2026-11-12},
	journal = {Adv. Phys. Res.},
	publisher = {John Wiley \& Sons, Ltd},
	author = {Geilen, Moritz and Verba, Roman and Hamadeh, Abbass and Nicoloiu, Alexandra and Narducci, Daniele and Dinescu, Adrian and Ender, Milan and Mohseni, Morteza and Ciubotaru, Florin and Weiler, Mathias and Müller, Alexandru and Hillebrands, Burkard and Adelmann, Christoph and Pirro, Philipp},
	month = jan,
	year = {2025},
	pages = {2400086},
}

@INPROCEEDINGS{Li_Hybrid-Magnon_2022,
  author={Li, Y. and Qian, J.-C. and Jiang, Z.-H. and Lo, T.-H. and Ding, D. and Draher, T. and Polakovic, T. and Pfaff, W. and Schleife, A. and Zuo, J.-M. and Kwok, W.-K. and Novosad, V. and Hoffmann, A.},
  booktitle={2022 International Electron Devices Meeting (IEDM)}, 
  title={Hybrid-Magnon Quantum Devices: Strategies and Approaches}, 
  year={2022},
  volume={},
  number={},
  pages={14.6.1-14.6.4},
  doi={10.1109/IEDM45625.2022.10019460}}

@ARTICLE{Awschalom_quantum_2021,
  author={Awschalom, David D. and Du, Chunhui Rita and He, Rui and Heremans, F. Joseph and Hoffmann, Axel and Hou, Justin and Kurebayashi, Hidekazu and Li, Yi and Liu, Luqiao and Novosad, Valentine and Sklenar, Joseph and Sullivan, Sean E. and Sun, Dali and Tang, Hong and Tyberkevych, Vasyl and Trevillian, Cody and Tsen, Adam W. and Weiss, Leah R. and Zhang, Wei and Zhang, Xufeng and Zhao, Liuyan and Zollitsch, Ch. W.},
  journal={IEEE Trans. Quantum Eng.}, 
  title={Quantum Engineering With Hybrid Magnonic Systems and Materials (Invited Paper)}, 
  year={2021},
  volume={2},
  number={},
  pages={1-36},
  doi={10.1109/TQE.2021.3057799}}

@article{lim_tracking_2020,
	title = {Tracking the {Suhl} instability versus angle and frequency for the backward volume mode in an yttrium iron garnet film},
	volume = {501},
	issn = {0304-8853},
	url = {https://www.sciencedirect.com/science/article/pii/S0304885319331440},
	doi = {10.1016/j.jmmm.2020.166441},
	urldate = {2026-11-13},
	journal = {J. Magn. Magn. Mater.},
	author = {Lim, Jinho and Bang, Wonbae and Trossman, Jonathan and Tsai, C. C. and Ketterson, John B.},
	month = may,
	year = {2020},
	pages = {166441},
}

@ARTICLE{mohseni_amplification_2023,
  author={Mohseni, Morteza and Hamadeh, Alexandre Abbass and Geilen, Moritz and Pirro, Philipp},
  journal={IEEE Transactions on Nanotechnology}, 
  title={Amplification and Frequency Conversion of Spin Waves Using Acoustic Waves}, 
  year={2023},
  volume={22},
  number={},
  pages={806-810},
  doi={10.1109/TNANO.2023.3331586}}

@article{lisenkov_magnetoelastic_2019,
  title = {Magnetoelastic parametric instabilities of localized spin waves induced by traveling elastic waves},
  author = {Lisenkov, Ivan and Jander, Albrecht and Dhagat, Pallavi},
  journal = {Phys. Rev. B},
  volume = {99},
  issue = {18},
  pages = {184433},
  numpages = {8},
  year = {2019},
  month = {May},
  publisher = {American Physical Society},
  doi = {10.1103/PhysRevB.99.184433},
  url = {https://link.aps.org/doi/10.1103/PhysRevB.99.184433}
}

@article{rückriegel_magnetoelastic_2014,
  title = {Magnetoelastic modes and lifetime of magnons in thin yttrium iron garnet films},
  author = {R{\"u}ckriegel, Andreas and Kopietz, Peter and Bozhko, Dmytro A. and Serga, Alexander A. and Hillebrands, Burkard},
  journal = {Phys. Rev. B},
  volume = {89},
  issue = {18},
  pages = {184413},
  numpages = {14},
  year = {2014},
  month = {May},
  publisher = {American Physical Society},
  doi = {10.1103/PhysRevB.89.184413},
  url = {https://link.aps.org/doi/10.1103/PhysRevB.89.184413}
}

@article{wang_room-temperature_2026,
	title = {A room-temperature cavity–magnonic source of correlated microwave magnon polariton pairs},
	volume = {9},
	copyright = {2026 The Author(s), under exclusive licence to Springer Nature Limited},
	issn = {2520-1131},
	url = {https://www.nature.com/articles/s41928-026-01689-y},
	doi = {10.1038/s41928-026-01689-y},
	language = {en},
	number = {9},
	urldate = {2026-11-18},
	journal = {Nat Electron},
	publisher = {Nature Publishing Group},
	author = {Wang, Qiuyuan and Karthigeyan, Aravind and Chou, Chung-Tao and Liu, Luqiao},
	month = sep,
	year = {2026},
	pages = {988--996},
}

\clearpage
\appendix 

\widetext
\begin{center}
    \LARGE \textbf{Supplemental Material}
\end{center}
\vspace{1em}

\setcounter{equation}{0}
\setcounter{figure}{0}
\setcounter{table}{0}
\setcounter{section}{0}
\renewcommand{\theequation}{S\arabic{equation}}
\renewcommand{\thefigure}{S\arabic{figure}}
\renewcommand{\thetable}{S\arabic{table}}
\renewcommand{\thesection}{S\arabic{section}}

\section{Magnon continuum mode dynamics in a thin film}

In this section we provide the magnon mode functions, nonlinear coupling rate, and dynamical equation for the thin-film setup considered in the main text. This section follows closely the derivations made in our previous work~\cite{Bruehlmann_2026_classical}.

The classical dynamics of the total magnetization field $\textbf{M}(\textbf{r},t)$ is governed by the Landau-Lifshitz Gilbert (LLG) equation~\cite{prabhakar2009spin,landau1980statisticalII}
\begin{align}
\begin{split}\label{SMeq:LLG}
    \frac{d}{dt} &\textbf{M}(\textbf{r},t) = \gamma_g \mu_0 \textbf{M}(\textbf{r},t) \times \textbf{H}_{\mathrm{eff}}(\mathbf{M},\textbf{r},t) + \frac{\lambda}{M_S}(\gamma_g \mu_0) \mathbf{M}(\textbf{r},t)\times(\textbf{M}(\textbf{r},t) \times \textbf{H}_{\mathrm{eff}}(\mathbf{M},\textbf{r},t)).
\end{split}
\end{align}
The constants $\gamma_g$, $\mu_0$, and $M_S$ are the electron gyromagnetic ratio, the vacuum permeability, and the saturation magnetization, respectively. The second term on the right-hand side leads to damping with a strength given by the  coefficient $\lambda$. In the system under consideration it is sufficient to include only the effects of exchange and dipole-dipole interaction in the effective field, $\textbf{H}_{\mathrm{eff}}(\mathbf{r},t) = H_0 \textbf{e}_z + \textbf{H}_x(\mathbf{r},t) +  \textbf{H}_{\mathrm{dm}}(\mathbf{r},t)$ with $\textbf{H}_x(\textbf{M},\textbf{r},t) =  \frac{A_x}{\mu_0 M_S^2} \nabla^2 \textbf{M}(\textbf{r},t)$ and $\textbf{H}_{\text{dm}}(\textbf{M},\textbf{r},t) = \frac{1}{4\pi} \nabla \int d^3 \textbf{r}^\prime \, \frac{\nabla_{\textbf{r}^\prime} \textbf{M}(\textbf{r}^\prime,t) }{|\textbf{r}- \textbf{r}^\prime|}$. Here we have introduced the uniform external field $H_0\equiv B_0/\mu_0$ and the exchange stiffness $A_x$. 
We undertake the spin-wave approximation, i.e., we seek solutions of the form 
\begin{align}\label{SMeq:magnetization}
    \mathbf{M}(\mathbf{r},t)=(M_S-\delta m(\mathbf{r},t)) \mathbf{e}_z+ \mathbf{m}(\mathbf{r},t),
\end{align} 
describing small magnetization oscillations $\vert\mathbf{m}\vert\ll M_S$ in the $x,y$ plane above the fully magnetized state. The correction $\delta m \approx (\mathbf{m}\cdot\mathbf{m})/(2M_S)$ ensures that the modulus $|\mathbf{M}|$ remains constant to second order in $\mathbf{m}/M_S$. \par

In the thin-film limit magnons are well described by the lowest band~\cite{kalinikos_theory_1986,CGB_2022}, and the mode decomposition of the magnetization reads
\begin{align}\label{SMeq:magnetization_decomposition}
    \mathbf{m}(\mathbf{r},t) = \int d^2\kvec s_\kvec(t) \mathbf{m}_\kvec(\mathbf{r}) + c.c. 
\end{align}
where $s_\kvec(t)$ and $\mathbf{m}_\kvec(\mathbf{r})$ are the amplitude and mode function of modes labeled by an in-plane wave vector $\kvec$. We choose as our mode decomposition the eigenmodes of the lossless, linearized LLG equation,
\begin{align}\label{LLGlinearized}
    \frac{d}{dt} \mathbf{m}(\mathbf{r},t) =\mathcal{D}_0 [ \mathbf{m}(\mathbf{r},t) ] - \lambda\mathbf{e}_z \times \mathcal{D}_0 [ \mathbf{m}(\mathbf{r},t) ],
\end{align}
with the linear integro-differential operator
\begin{equation}
    \mathcal{D}_0 [ \mathbf{m}(\mathbf{r},t) ] \equiv \mathbf{e}_z \times \left(\omega_H - \omega_M \frac{A_x}{\mu_0 M_S^2} \nabla^2\right) \mathbf{m}(\mathbf{r},t) 
    - \omega_M \mathbf{e}_z \times \frac{1}{4\pi} \nabla \int d^3 \textbf{r}^\prime \, \frac{\nabla_{\textbf{r}^\prime} \textbf{m}(\textbf{r}^\prime,t) }{|\textbf{r}- \textbf{r}^\prime|},
\end{equation}
using $\omega_M \equiv |\gamma_g|\mu_0 M_S$, and $\omega_H \equiv |\gamma_g|B_0$. The eigenmode solutions oscillate with frequency $\omega_\kvec$ and have a spatial mode function fulfilling $-i\omega_\kvec \mathbf{m}_\kvec(\mathbf{r}) =\mathcal{D}_0 [ \mathbf{m}_\kvec(\mathbf{r}) ]$, which is given by~\cite{kalinikos_theory_1986,CGB_2022}
\begin{gather}\label{SMeq:magnon_modefunction}
    \mathbf{m}_\kvec(\mathbf{r}) = \mathcal{N}_\kvec e^{i\mathbf{kr}_{\scriptscriptstyle{\parallel}}} \begin{pmatrix}
        \sqrt{\nu_y(\kvec)} \\ i \sqrt{\nu_x(\kvec)} \\ 0 
    \end{pmatrix}, 
\end{gather}
with
\begin{align}
    \nu_x(\kvec) &\equiv \frac{\omega_H}{\omega_M} + \frac{A_x}{\mu_0 M_S^2} k^2 - \frac{( e^{-k d } -1)}{k d}, \\
    \nu_y(\kvec) &\equiv \frac{\omega_H}{\omega_M} + \frac{A_x}{\mu_0 M_S^2} k^2 + \frac{k_y^2}{k^2} \left( 1 +  \frac{( e^{-k d } -1)}{k d}\right) .
\end{align}
The variable $\mathbf{r}_{\scriptscriptstyle{\parallel}} = (0,y,z)$ denotes an in-plane position vector and  $\mathcal{N}_\kvec = \sqrt{\omega_M/(2 d(2\pi)^2 \omega_\kvec)}$ is a normalization factor. The associated eigenmode frequencies are given by $\omega_\kvec = \omega_M \sqrt{\nu_x(\kvec)\nu_y(\kvec)}$. The eigenmode functions Eq.~\eqref{SMeq:magnon_modefunction} form an orthonormal basis ($\bracket{\mathbf{m}_\kvec}{\mathbf{m}_{\kvec'}}= i \, \delta(\kvec-\kvec^\prime) $) under the inner product $\bracket{\mathbf{g}}{\mathbf{f}} \equiv \int d\mathbf{r}\, (\mathbf{e}_z \times \mathbf{g}^*(\mathbf{r})) \cdot \mathbf{f}(\mathbf{r})$. The mode amplitudes $s_\kvec(t)$ are determined by projection of the magnetization field Eq.~\eqref{SMeq:magnetization_decomposition} onto the eigenmodes, i.e., $s_\kvec(t) = -i \bracket{\mathbf{m}_\kvec}{\mathbf{m}(\mathbf{r},t)}$. This results in the following differential equation for the magnon amplitudes
\begin{equation}\label{skeqlinear}
    \dot{s}_\kvec = (-i\omega_\kvec - \gamma_\kvec/2)s_\kvec
\end{equation}
with the damping rate  $\gamma_\kvec= \lambda \omega_M (\nu_x(\kvec)+\nu_y(\kvec)) $. \par

Nonlinear dynamics arise after inserting the magnetization Eq.~\eqref{SMeq:magnetization} into the full LLG equation~\eqref{SMeq:LLG} and applying the mode decomposition \eqref{SMeq:magnetization_decomposition}. Projecting onto a specific eigenmode $\kvec$ using $\bra{\mathbf{m}_\kvec}$ and keeping terms up to second order in $\mathbf{m}/M_S$ we obtain the dynamical equations for the amplitudes including three-magnon scattering (3MS),
\begin{align}\label{SMeq:magnon_eom_intermediate}
    \dot s_\kvec = \left(-i \omega_\kvec -\frac{\gamma_\kvec^m}{2}\right) s_\kvec -i\iint d^2\qvec_1 d^2\qvec_2 \mspace{-4mu}  \sum_{\alpha,\beta=\pm}  \bracket{\mathbf{m}_\kvec}{D^{(3)}[\mathbf{m}_{\qvec_1}^\alpha,\mathbf{m}_{\qvec_2}^\beta]} s_{\qvec_1}^\alpha s_{\qvec_2}^\beta  
\end{align}
where for compactness we define variables and their complex conjugates with a $\pm$ index, i.e., $\mathbf{m}_{k}^+(\mathbf{r})\equiv\mathbf{m}_{k}(\mathbf{r})$, $\mathbf{m}_{k}^-(\mathbf{r})\equiv\mathbf{m}_{k}^*(\mathbf{r})$, $s_{\kvec}^+\equiv s_{\kvec}$, and $s_{\kvec}^-\equiv s_{\kvec}^*$, and the vector
\begin{align}
    D^{(3)}[\mathbf{m}_{\qvec_1},\mathbf{m}_{\qvec_2}] \equiv \frac{\omega_M}{2 M_S}  
    \biggl[ \mathbf{e}_z \times \Bigl(\int d\mathbf{r}^\prime \mathcal{\bar{\bar G}}(\mathbf{r\mspace{-4mu}-\mspace{-4mu}r}^\prime) \mathbf{e}_z \bigl( \mathbf{m}_{\qvec_1}(\mathbf{r}^\prime) \cdot \mathbf{m}_{\qvec_2}(\mathbf{r}^\prime)\bigr)\Bigr)
    - 2\mathbf{m}_{\qvec_1}(\mathbf{r}) \times \Bigl( \int d\mathbf{r}^\prime \mathcal{\bar{\bar G}}(\mathbf{r - r}^\prime) \mathbf{m}_{\qvec_2}(\mathbf{r}^\prime) \Bigr) \biggr].
\end{align}
Here we have introduced for the demagnetizing field a Green's function notation $\textbf{H}_{\text{dm}}(\textbf{M},\textbf{r},t)  = \int d\mathbf{r}^\prime \mathcal{\bar{\bar G}}(\mathbf{r-r}^\prime) \mathbf{M}(\mathbf{r}^\prime,t)$, with a Green's function~\cite{CGB_2022} 
\begin{align}\label{SMeq:Greens_function}
    2\rttensor{\mathcal{G}}(\textbf{r}-\textbf{r}^\prime) \equiv \delta(\textbf{r}_{\scriptscriptstyle{\parallel}}-\textbf{r}_{\scriptscriptstyle{\parallel}}^\prime)\mathbf{e}_x \mathbf{e}_x^T -\int d^2 \textbf{q} \, \frac{e^{i\textbf{q} (\textbf{r}_{\scriptscriptstyle{\parallel}}-\textbf{r}_{\scriptscriptstyle{\parallel}}^\prime)}}{(2\pi)^2q} e^{-q|x-x^\prime|} \mathbf{v}\mathbf{v}^T,
\end{align}
and with the vectors $\mathbf{e}_x=(1,0,0)$ and $\mathbf{v}\equiv(-iq\,\text{sign}(x-x'),q_y,q_z)$. Applying a rotating wave approximation, i.e., neglecting the rapidly oscillating term proportional to $s^* s^*$, the classical equation of motion takes the form 
\begin{align}\label{SMeq:magnon_eom}
    \dot s_\kvec = \left( -i\omega^\text{m}_\kvec-\frac{\gamma^\text{m}_{\kvec}}{2} \right) s_\kvec 
    +\int d^2\qvec\; \Bigl( -ig_{\text{cl}}(\qvec,\kvec-\qvec,\kvec) \, s_{\qvec} s_{\kvec-\qvec} -i2g_{\text{cl}}^*(\kvec,\qvec,\kvec+\qvec) \,  s_{\qvec}^* s_{\kvec+\qvec} \Bigr)
\end{align}
with the 3MS coupling rate density 
\begin{equation}
    g_{\text{cl}}(\qvec,\kvec-\qvec,\kvec) \equiv -i \frac{d(2\pi)^2\omega_M}{2 M_S} \mathcal{N}_{\kvec} \mathcal{N}_{\qvec} \mathcal{N}_{\kvec-\qvec}
    \left( J^-_{\qvec,\kvec-\qvec,\kvec}-J^+_{\kvec,\qvec,\kvec-\qvec}-J^+_{\kvec,\kvec-\qvec,\qvec}\right) 
\end{equation}
where we define the functions
\begin{align}
    J^\pm_{\qvec_1,\qvec_2,\qvec_3}\equiv\left( \pm\sqrt{\nu_x(\qvec_1) \nu_x(\qvec_2)} + \sqrt{\nu_y(\qvec_1) \nu_y(\qvec_2)} \right) 
    \sqrt{\nu_x(\qvec_3)}\biggl( -1 +\frac{1}{q_3d} (1-e^{-q_3d}) \biggr) \frac{q_{3y} q_{3z}}{q_3^2}.
\end{align}
The first coupling term of Eq.~\eqref{SMeq:magnon_eom} describes a magnon combination interaction ($\{\qvec,\kvec-\qvec\}\to\kvec$) while the second term describes a magnon scattering interaction of modes $\kvec+\qvec\to\{\kvec,\qvec\}$. \par

This derivation of the nonlinear magnon dynamical equation can be extended to also include four-magnon scattering interactions (4MS). The interaction terms and coupling rates can be found in our previous work~\cite{Bruehlmann_2026_classical}. For the driven mode considered in Fig.~2a of the main text, the threshold drive amplitude $\alpha_\text{th}$ for parametric amplification via 3MS is lower than the one for 4MS by a factor $~10^{-3}$. It is therefore justifiable to neglect 4MS coupling effects for the system parameters applied here. \par

The quantum version of equation~\eqref{SMeq:magnon_eom} is obtained by canonical quantization, i.e., by promoting magnon mode amplitudes to ladder operators (see e.g., Refs.~\cite{mills_quantum2006,gonzalez-ballestero_quantum_2020,gonzalez-ballestero_theory_2020} and references therein) as $s_\mathbf{k} \rightarrow \mathcal{M}_0 \hat s_\mathbf{k}$ with zero-point magnetization $\mathcal{M}_0 \equiv  \sqrt{\hbar\vert \gamma_g\vert M_S}$, and imposing canonical bosonic commutation relations $\left[\hat{s}_\mathbf{k},\hat{s}_\mathbf{q}\right]=0$ and $\left[\hat{s}_\mathbf{k},\hat{s}_\mathbf{q}^\dagger\right]=\delta(\mathbf{k}-\mathbf{q})$. The coherent dynamics of Eq.~\eqref{SMeq:magnon_eom} become Heisenberg picture equations, which are generated by the Hamiltonian
\begin{equation}\label{SMeq:continuum_Hamiltonian}
    \Hop = \int d\kvec\, \hbar\omega_\kvec \, \sop_\kvec^\dagger \sop_\kvec + \iint d\kvec\, d\kvec^\prime \hbar \left( g(\kvec,\kvec^\prime,\kvec+\kvec^\prime)\sop_\kvec \sop_{\kvec^\prime} \sop_{\kvec+\kvec^\prime}^\dagger +h.c. \right) 
\end{equation}
where $g= \mathcal{M}_0 g_{\text{cl}}$. The Gilbert damping is modeled by the addition of thermal dissipators to the quantum master equation
\begin{align}\label{SMeq:master_eq_preTransform}
    \frac{d}{dt} \hat{\rho} = -\frac{i}{\hbar} [\Hop, \hat{\rho}] + \int d\kvec \frac{\gamma_\kvec}{2} \Bigl( (1+\bar{n}_\kvec) \mathcal{L}_{\sop_\kvec} + \bar{n}_\kvec \mathcal{L}_{\sop_\kvec^\dagger} \Bigr).
\end{align}
Here $\hat{\rho}$ is the density matrix, $\bar{n}_\kvec = (\exp(\hbar \omega_\kvec/k_B T)-1)^{-1}$ the thermal occupation of a bath at temperature $T$ which acts as a source for the Gilbert damping. The Lindblad operators take the form $\mathcal{L}_{\sop_\kvec} = \sop_\kvec \hat{\rho} \sop_\kvec^\dagger - \{\sop_\kvec^\dagger \sop_\kvec , \hat{\rho}\}/2$.

\section{Magnon dynamics in the collective mode basis}

In the main text we have used a discrete set of collective magnon modes $\bTop_j$ centered around wavevectors $\kvec_j$ for $j \in \{0,1,2\}$. In this section we define a full basis of magnon modes which incorporates these collective operators and show how it is related to the continuum modes $\kvec$. We show why the system at hand can be well approximated by only the subspace of the collective modes $\bTop_j$ and calculate the analytical expressions for the coupling rate $G$ and driving strength coefficient $\alpha_\text{d}$. Furthermore we introduce the mode filter function $f_j$ used in numerical calculations of the main text. \par

We consider the magnon dynamics of a thin film ferromagnet as described by the quantum master equation~\eqref{SMeq:master_eq_preTransform}. In the parameter regime of the coherently driven magnon system described in the main text where parametric down-conversion by 3MS is observable only three modes $\{\kvec_0, \kvec_1, \kvec_2 \}$, where $\kvec_2 = \kvec_0-\kvec_1$, become significantly populated~\cite{boardman_three_1988,mohseni_controlling_2021,ordonez-romero_thee_2009,Bruehlmann_2026_classical,Hwang_2026_Harmonic}. The finite linewidth of both the antenna-driven pump and the parametrically amplified modes naturally suggests a description in terms of a set of collective modes rather than the full continuum of plane-wave modes. We define three collective mode operators $\bTop_j$ representing a spectrally localized superposition of continuum magnon modes centered around $\kvec_j$ and an infinite set of delocalized operators $\{\Bop_\alpha\}$ to complete the basis.
\begin{align}
    \bTop_j &= \int d\kvec \, f_j(\kvec) \sop_\kvec, \qquad j=0,1,2\label{SMeq:b_operator}\\
    \Bop_\alpha &= \int d\kvec \, D_\alpha(\kvec) \sop_\kvec , \qquad \alpha \in \mathbb{N}
\end{align}
The orthonormal set of square-normalized functions $\{f_0,f_1,f_2,\{D_\alpha\}\}$ must span the one-particle Hilbert space $L^2(\mathbb{R}^2)$ of square integrable functions. The new discrete operators thus satisfy bosonic commutation relations $[ \bTop_i , \bTop_j ] =0$, $[ \bTop_i , \bTop_j^\dagger ] =\delta_{i,j}$, $[ \Bop_\alpha , \Bop_\beta ] = 0$, $[ \Bop_\alpha , \Bop_\beta^\dagger ] = \delta_{\alpha,\beta}$, $[ \bTop_j , \Bop_\beta^\dagger ] = 0$, and $[ \bTop_j , \Bop_\beta ] = 0$. Under this definition $\sop_\kvec$ can be decomposed in the new basis by $\sop_\kvec = \sum_j f_j^*(\kvec) \bTop_j + \sum_\alpha D_\alpha^*(\kvec) \Bop_\alpha$ which allows us to calculate the equations of motion of the collective mode operators
\begin{multline}
    \frac{d}{dt} \bTop_j = -i \omega_{\kvec_j} \bTop_j \\+ \sum_{m,n} \iint d\qvec \, d\kvec \, \bigl( -i g(\qvec , \kvec - \qvec , \kvec) f_j(\kvec) f_m^*(\qvec) f_n^*(\kvec-\qvec) \bTop_n \bTop_m - i 2 g^*(\kvec , \qvec , \kvec + \qvec) f_j(\kvec) f_m^*(\kvec+\qvec) f_n(\qvec) \bTop_n^\dagger \bTop_m \bigr)\\
    + \sum_{m,\alpha} \iint d\qvec \, d\kvec \, \bigl( -i g(\qvec , \kvec - \qvec , \kvec) f_j(\kvec) f_m^*(\qvec) D_\alpha^*(\kvec-\qvec) \Bop_\alpha \bTop_m - i 2 g^*(\kvec , \qvec , \kvec + \qvec) f_j(\kvec) f_m^*(\kvec+\qvec) D_\alpha(\qvec) \Bop_\alpha^\dagger \bTop_m \bigr)\\
    + \sum_{\alpha,m} \iint d\qvec \, d\kvec \, \bigl( -i g(\qvec , \kvec - \qvec , \kvec) f_j(\kvec) D_\alpha^*(\qvec) f_m^*(\kvec-\qvec) \bTop_m \Bop_\alpha - i 2 g^*(\kvec , \qvec , \kvec + \qvec) f_j(\kvec) D_\alpha^*(\kvec+\qvec) f_m(\qvec) \bTop_m^\dagger \Bop_\alpha \bigr)\\
     + \sum_{\alpha,\beta} \iint d\qvec \, d\kvec \, \bigl( -i g(\qvec , \kvec - \qvec , \kvec) f_j(\kvec) D_\alpha^*(\qvec) D_\beta^*(\kvec-\qvec) \Bop_\beta \Bop_\alpha - i 2 g^*(\kvec , \qvec , \kvec + \qvec) f_j(\kvec) D_\alpha^*(\kvec+\qvec) D_\beta(\qvec) \Bop_\beta^\dagger \Bop_\alpha \bigr)
\end{multline}
The coupling of collective modes is only significant if the associated functions ($f_j$, $D_\alpha$) overlap strongly around triplets of conserved momenta (e.g., $\{ \kvec, \qvec, \kvec+\qvec \}$), i.e., interactions between the collective modes narrowly localized around momentum conserving wavevectors are much stronger than the other couplings. We thus retain only those significant couplings and find the closed system of equations
\begin{align}
    \frac{d}{dt} \bTop_0 &= -i\omega_{\kvec_0} \bTop_0 - i G \bTop_1 \bTop_2  \label{SMeq:EOM_0}\\
    \frac{d}{dt} \bTop_1 &= -i\omega_{\kvec_1} \bTop_1 - i G^* \bTop_2^\dagger \bTop_0 \label{SMeq:EOM_1}\\
    \frac{d}{dt} \bTop_2 &= -i\omega_{\kvec_2} \bTop_2 - i G^* \bTop_1^\dagger \bTop_0 \label{SMeq:EOM_2}
\end{align}
with coupling rate $G \equiv 2\,c\,g(\kvec_1,\kvec_2,\kvec_0)$ where the coefficient $c$ is defined by the equation
\begin{align}
    \label{SMeq:collective_coeff_1}
    \iint d\kvec\,d\qvec\, f_j(\kvec) f_i^*(\qvec) f_l^*(\kvec-\qvec) = c \, \delta_{j,0} ( \delta_{i,1} \delta_{l,2} + \delta_{i,2} \delta_{l,1} ) .
\end{align}
We also plug the decomposition of $\sop_\kvec$ in the new basis into the thermal dissipation terms. Over the width of envelopes $f_j(\kvec)$ the magnon decay rate $\gamma_\kvec$ and the bath thermal occupation $\bar{n}_\kvec$ vary only negligibly. Therefore we can use the orthonormality of $\{f_j,D_\alpha\}$ to write
\begin{multline}
    \label{SMeq:collective_mode_dissipators}
    \int d\kvec \frac{\gamma_\kvec}{2} \Bigl( (1+\bar{n}_\kvec) \mathcal{L}_{\sop_\kvec} + \bar{n}_\kvec \mathcal{L}_{\sop_\kvec^\dagger} \Bigr) = \sum_j \frac{\gamma_j}{2} \Bigl( (1+\bar{n}_j) \mathcal{L}_{\bTop_j} + \bar{n}_j \mathcal{L}_{\bTop_j^\dagger} \Bigr)\\
    + \sum_{\alpha,\beta} \int d\kvec\, \frac{\gamma_\kvec}{2} D_\alpha^*(\kvec) D_\beta(\kvec) \Bigl( (1+\bar{n}_\kvec) \mathcal{L}_{\Bop_\alpha,\Bop_\beta^\dagger} + \bar{n}_\kvec \mathcal{L}_{\Bop_\beta^\dagger,\Bop_\alpha} \Bigr)
\end{multline}
with $\gamma_j=\gamma_{\kvec_j}$ and $\bar{n}_j = \bar{n}_{\kvec_j}$ for $j\in\{0,1,2\}$, and Lindblad operators
\begin{align}
    \mathcal{L}_{\bTop_j} &= \bTop_j \hat{\rho} \bTop_j^\dagger -\{\bTop_j^\dagger \bTop_j, \hat{\rho} \}/2 , \\
    \mathcal{L}_{\Bop_\alpha,\Bop_\beta^\dagger} &= \Bop_\alpha \hat{\rho} \Bop_\beta^\dagger -\{\Bop_\beta^\dagger \Bop_\alpha, \hat{\rho} \}/2 .
\end{align}
For the closed system of $\{\bTop_j\}$ (Eqs.~\eqref{SMeq:EOM_0}-\eqref{SMeq:EOM_2}) only the first line of equation~\eqref{SMeq:collective_mode_dissipators} is relevant. \par

For numerical calculations we choose $f_j, \; j\in\{0,1,2\}$ to take the form of square-normalized anisotropic Gaussians:
\begin{align}
    f_j(\qvec) = \frac{\exp\left( -\frac{(q_y - k_{j,y})^2}{4\sigma_{j,y}^2} -\frac{(q_z - k_{j,z})^2}{4\sigma_{j,z}^2}\right)}{ \sqrt{2 \pi \sigma_{j,y} \sigma_{j,z}} }
\end{align}
If the centers $\kvec_j$ are well separated with regards to the widths $\sigma_j$, which is always true in the examples considered in the main text, then these functions are effectively orthonormal. Otherwise one may use Gram–Schmidt orthonormalization~\cite{barnett_methods_1997}. We choose the width of the collective modes such that the associated spectral variance is given by the intrinsic linewidth of the carrier mode $\Delta \omega_j \sim \gamma_{\kvec_j}$. Spectral and phase-space variance are related by Taylor expansion of the dispersion relation around the carrier mode $\kvec_j$. We thus use  
\begin{align}
   \sigma_{j,\alpha} = \Delta k_\alpha = \frac{ -v_{g,\alpha} \pm \sqrt{ v_{g,\alpha}^2 + 2 \gamma_{\kvec_j} \partial_\alpha^2 \omega (\kvec_j) } }{\partial_\alpha^2 \omega (\kvec_j) } 
\end{align}
for $\alpha\in\{y,z\}$, where $v_{g,\alpha}$ is the group velocity in direction $\alpha$. The inclusion of second-order terms is necessary since $v_{g,\alpha}$ can be zero in the nonlinear dispersion of magnons. The operator basis can be completed by, e.g., projecting a set of 2-dimensional Hermite functions~\cite{Celeghini_2021_Hermite} onto the orthogonal complement space of $\{f_j\}$. We, however, have no need for an explicit expression here. With this choice of $f_j$ the coefficient appearing in the definition of the collective mode coupling rate $G$ takes the form
\begin{align}
    c = \mathcal{N}_0 \mathcal{N}_1 \mathcal{N}_2 \frac{4\pi}{\sqrt{ \sigma_{0,y}^2 \sigma_{1,y}^2 + \sigma_{0,y}^2 \sigma_{2,y}^2 + \sigma_{1,y}^2 \sigma_{2,y}^2 }} \frac{4\pi}{\sqrt{ \sigma_{0,z}^2 \sigma_{1,z}^2 + \sigma_{0,z}^2 \sigma_{2,z}^2 + \sigma_{1,z}^2 \sigma_{2,z}^2 }}
\end{align}
with normalization factors $\mathcal{N}_j= (2\pi \sigma_{j,y} \sigma_{j,z})^{-\frac{1}{2}}$.\par

We furthermore introduce a drive to the system through the Hamiltonian $\Hop_\text{drive} = \int dV\, \hat{\mathbf{M}} \cdot \mathbf{B}_\text{drive}$ where we apply a classical in-plane polarized ($\mathbf{e}_\text{d}\cdot \mathbf{e}_z=0$) drive field of strength $B_\text{d}$
\begin{align}
    \mathbf{B}_\text{drive} = \frac{B_\text{d} \mathbf{e}_\text{d}}{2\sqrt{2\pi \sigma_{0,y} \sigma_{0,z}}} \int d\kvec\, (f_0(\kvec) e^{-i\omega_{\kvec_0}t +i\kvec \mathbf{r}_{\scriptscriptstyle{\parallel}}} + cc. ).
\end{align}
centered around---and oscillating coherently with---$\kvec_0$. Under the assumption that the mode function changes slowly compared to the width of $f_0$ the coherent drive Hamiltonian takes the form
\begin{align}
    \Hop_\text{drive} = \hbar ( i \alpha_\text{d} e^{-i\omega_{\kvec_0} t} \bTop_0^\dagger + h.c. )
\end{align}
with $\alpha_\text{d} \equiv -i (2\pi)^2 d \mathcal{M}_0 (\mathbf{m}_{\kvec_0}^* \cdot \mathbf{e}_\text{d}) B_\text{d} / (\hbar\, 2\sqrt{2\pi \sigma_{0,y} \sigma_{0,z}} )$. The full Hamiltonian representing the magnon dynamics is thus
\begin{align}\label{SMeq:magnon_hamiltonian_original}
    \Hop_\text{m} = \sum_{j=0}^2 \hbar \omega_{\kvec_j} \bTop_j^\dagger \bTop_j &+ \hbar G \, \bTop_1 \bTop_2 \bTop_0^\dagger + \hbar G^* \, \bTop_0 \bTop_2^\dagger \bTop_1^\dagger + \hbar (i \alpha_\text{d} e^{-i\omega_{\kvec_0} t} \bTop_0^\dagger + h.c. ) .
\end{align}

\section{Magnon-microwave coupling}\label{sec:SM_magnon-microwave}
In this section we model the inductive coupling of magnons to transmission line microwaves through antennas. We show the coupling rate expressions and discuss constraints on the coupling strengths. The interaction between magnetization $\hat{\mathbf{M}}(\rvec)$, and thus magnon modes $\bTop_j$, and the magnetic field $\hat{\mathbf{h}}(\rvec)$ of the antennas is given by the Zeeman-type coupling
\begin{align}\label{SMeq:detector_int1}
    \Hop_{\text{int}} &= -\mu_0 \int d^3\rvec \, \hat{\mathbf{h}}(\rvec) \cdot \hat{\mathbf{M}}(\rvec) .
\end{align}
We consider two structured antennas ($j=1,2$) located at positions $\rvec_j$ and phase-engineered to couple to modes with wavevector $\qvec_j$. The associated magnetic field in terms of bosonic transmission line microwave mode operators $\aop_\omega^{(j)}$ takes the form~\cite{gardiner_quantum_2004,collett_input_1984,ciubotaru_all_2016}
\begin{align}\label{SMeq:detector_hfield}
    \hat{\mathbf{h}}(\rvec) = \sum_{j=1}^2 \left( \mathbf{e}_{j} F_\text{D}(\rvec-\rvec_{j}) e^{i \qvec_j (\rvec-\rvec_{j})} \int d\omega\, \sqrt{K_j(\omega)} \aTop_\omega^{(j)} + h.c. \right)
\end{align}
The $h$-field polarization ($\mathbf{e}_{j}$) must be in the $y,z$-plane in order to achieve coherent coupling to the magnetization, with the coupling strength controlled by $K_j(\omega)$. The real-valued filter function $F_\text{D}$ describes the localization of the detector around the center $\rvec_{j}$ in real space, while the exponential term ensures localization in phase space is centered around wavevector $\qvec_j$. We consider antennas with the Fourier transform $F_\text{D}(\qvec_j-\kvec) = \int d^2\rvec \, F_\text{D}(\rvec-\rvec_{j}) e^{-i\kvec \rvec} e^{i \qvec_j (\rvec-\rvec_{j})}$ sufficiently narrow and centered around magnon modes ($\qvec_j=\kvec_j$), e.g., $F_\text{D}(\qvec_j-\kvec)=f_j(\kvec)$, which produce a beamsplitter interaction Hamiltonian 
\begin{align}\label{SMeq:interaction_Hamiltonian}
    \Hop_{\text{int}} &= i\hbar \sum_{j=1}^2 \int d\omega\, \sqrt{\kappa_{j}(\omega)} \aTop_\omega^{(j)} \bTop_j^\dagger + h.c.
\end{align}
with coupling rate
\begin{align}\label{SMeq:kappa}
    \sqrt{\kappa_{j}(\omega)} = -\frac{\mu_0 d \mathcal{M}_0}{i\hbar}\sqrt{K_j(\omega)} \biggl( \int d^2\kvec\, (\mathbf{e}_{j}\cdot \mathbf{m}_\kvec^*) F_\text{D}(\qvec_j-\kvec) f_j(\kvec) \Bigr) .
\end{align}
Note that each detector only significantly couples to one magnon operator $\bTop_j$. Coupling to magnon modes $\Bop_\alpha$ is much weaker and thus neglected, due to $D_\alpha(\kvec)$ being delocalized. In this work, for the formulation of the input-output relation we require the interaction to be weak enough such that the magnon modes act as a Markovian bath for the microwave photons. This means the bath correlation time is much shorter than the system relaxation time, which translates to a frequency-independent coupling density $\sqrt{\kappa_{j}(\omega)}\equiv \sqrt{\kappa_{j}}$ over the relevant bandwidth. This assumption also ensures that photons do not significantly affect the parametric amplification mode selection process.

\section{Classical mean-fields in parametric down-conversion}
In this section we present the solution of classical magnon and microwave mean-field in the long time limit. This section builds on the calculations in Ref.~\cite{Bruehlmann_2026_classical}. The dynamics of the full magnon-microwave system are governed by the quantum master equation
\begin{align}\label{SMeq:master_eq_collective_preTransform}
    \frac{d}{dt} \hat{\rho} = -\frac{i}{\hbar} [\Hop, \hat{\rho}] + \sum_j \frac{\gamma_j}{2} \Bigl( (1+\bar{n}_j) \mathcal{L}_{\bTop_j} + \bar{n}_j \mathcal{L}_{\bTop_j^\dagger} \Bigr)
\end{align}
where $\rho$ is the density matrix and thermal magnon dissipation is described by decay rates $\gamma_j$, thermal bath occupation numbers $\bar{n}_j$ and Lindblad operators $\mathcal{L}_{\bTop_j} = \bTop_j \hat{\rho} \bTop_j^\dagger - \{\bTop_j^\dagger \bTop_j , \hat{\rho}\}/2$. The full Hamiltonian $\Hop = \Hop_\text{m} + \Hop_\text{mw} + \Hop_\text{int}$---where $\Hop_\text{m}$ is the magnon Hamiltonian (Eq.~\eqref{SMeq:magnon_hamiltonian_original}, $\Hop_\text{mw} = \sum_j \int d\omega \, \hbar \omega \aTop_\omega^{(j)\dagger} \aTop_\omega^{(j)}$ is the photon Hamiltonian, and $\Hop_{\text{int}}$ the interaction Hamiltonian (Eq.~\eqref{SMeq:interaction_Hamiltonian})---is time dependent. To treat the long time limit we look at classical dynamics of this nonlinear system in a rotating frame, i.e., the dynamics of expectation values $\beta_j \equiv\langle \bTop_j \rangle e^{-i\Omega_j t}$ ($j\in\{0,1,2\}$) and $\alpha_j(\omega) \equiv\langle \aTop_\omega^{(j)}(t) \rangle e^{-i\Omega_j t}$ ($j\in\{1,2\}$) whose oscillation frequencies must satisfy $\Omega_0 = \omega_{\kvec_0}$, and $ \Omega_1 + \Omega_2 = \Omega_0$. We derive from the master equation the equations of motion for the classical amplitudes and neglect all correlations between the modes under a mean-field approximation, e.g., $\langle \bTop_j \bTop_k\rangle = \beta_j \beta_k e^{i(\Omega_j+\Omega_k)t}$. The steady state equations for the classical amplitudes are
\begin{gather}
    0 = (-i\Delta_0 -\frac{\gamma_{0}}{2}) \beta_0 - i G \beta_1 \beta_2 + \alpha_\text{d}\\
    0 = (-i\Delta_1 -\frac{\gamma_{1}}{2}) \beta_1 - i G^* \beta_2^* \beta_0 + \int_0^\infty d\omega\, \sqrt{\kappa_{1}(\omega)} \alpha_1(\omega) \\
    0 = (-i\Delta_2 -\frac{\gamma_{2}}{2}) \beta_2 - i G^* \beta_1^* \beta_0 + \int_0^\infty d\omega\, \sqrt{\kappa_{2}(\omega)} \alpha_2(\omega) \\
    0 = \int_0^\infty d\omega\, \left( \Delta_\omega^{(j)} \alpha_j^*(\omega) + i\sqrt{\kappa_{j}(\omega)} \beta_j^* \right)\qquad j\in\{1,2\} \label{SMeq:class_field_EOM_4}
\end{gather}
where $\Delta_j = \omega_{\kvec_j} - \Omega_j$ ($j\in\{0,1,2\}$) and $\Delta_\omega^{(j)}=\omega-\Omega_j$ ($j\in\{1,2\}$) are the oscillation frequencies in the rotating frame.\par

\begin{figure}
    \centering
    \includegraphics[width=.5\linewidth]{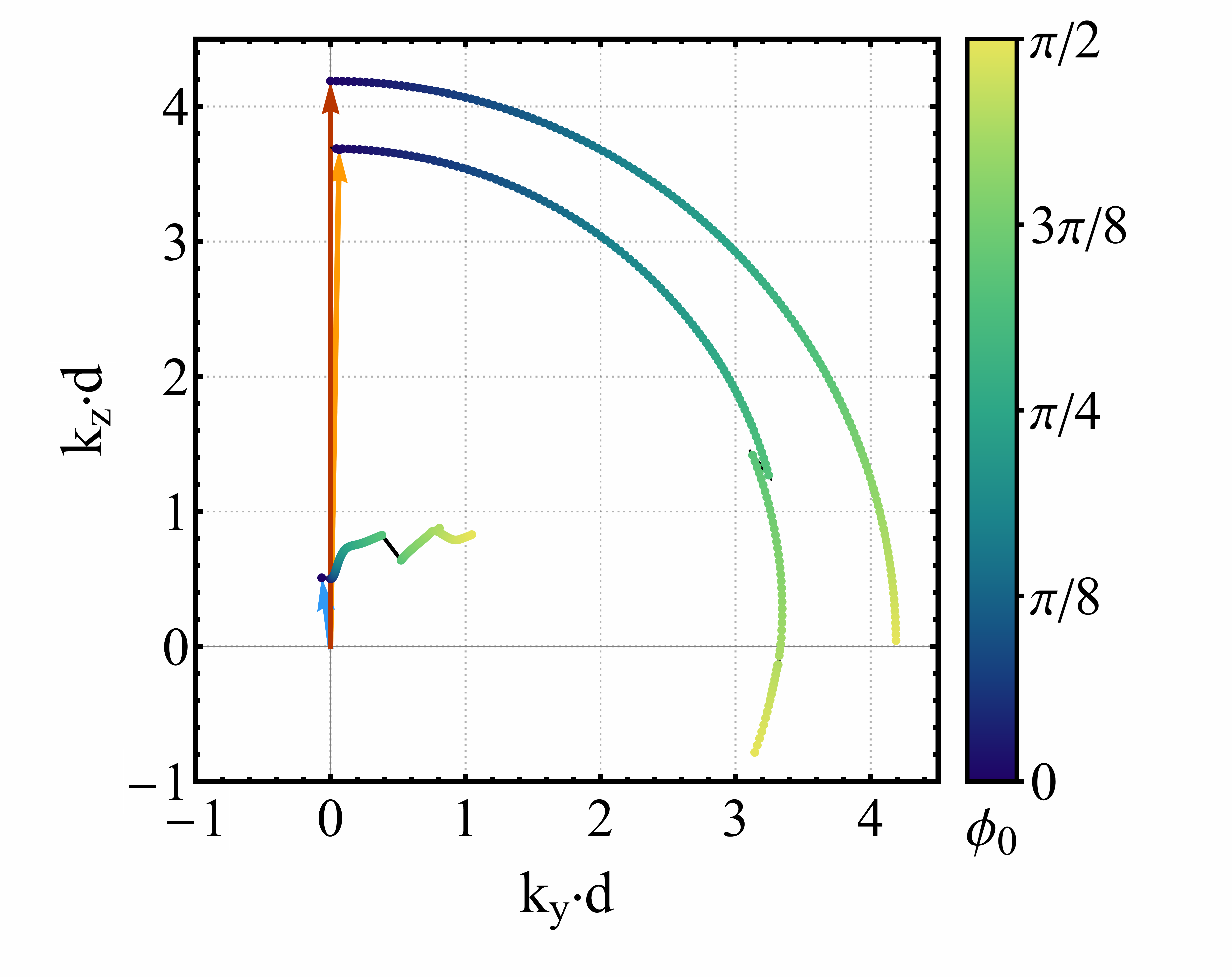}
    \caption{The directionality of the two amplified modes (orange and blue arrows) changes with the orientation of the external field $\mathbf{B}_0$ which is quantified by the angle $\phi_0$ between $\mathbf{B}_0$ and $\kvec_0$. The dots mark where the wavevectors point for the associated value of $\phi_0$. For reference we also show the directionality of the driven mode (red arrow) as a function of $\phi_0$. }
    \label{fig:directionality}
\end{figure}

As we have mentioned earlier, we consider weak enough magnon-microwave coupling such that microwaves do not affect the parametric amplifiaction process of magnons. In the decoupled case consistency of the steady-state equations demands~\cite{Bruehlmann_2026_classical} the following oscillation frequencies:
\begin{equation}
    \left[
    \begin{array}{c}
         \Omega_1  \\
         \Omega_2 
    \end{array}
    \right]=\frac{1}{\gamma_{1}+\gamma_{2}}\left[
    \begin{array}{c}
         \gamma_{2}\omega_{\kvec_1}-\gamma_{1}\omega_{\kvec_2}+\gamma_{1}\omega_{\kvec_0}  \\
         \gamma_{1}\omega_{\kvec_2}-\gamma_{2}\omega_{\kvec_1}+\gamma_{2}\omega_{\kvec_0} 
    \end{array}
    \right]
\end{equation}
Analogously to the derivation in our previous paper~\cite{Bruehlmann_2026_classical} we find that the solution of parametrically amplified magnon modes $\beta_{1,2}$ only exists for drive amplitudes $|\alpha_\text{d}|>|\alpha_\text{th}|$ above the threshold 
\begin{equation}
    |\alpha_{\text{th}}|^2 = \left| i\Delta_0 + \gamma_{0}/2 \right|^2  \frac{\left|(i\Delta_1 + \gamma_{1}/2)(i\Delta_2 + \gamma_{2}/2)\right|}{|G|^2}
\end{equation}
The parametrically excited modes are those with the lowest threshold value~\cite{suhl_theory_1957,zakharov_spin-wave_1975,Bruehlmann_2026_classical,Hwang_2026_Harmonic}. In the main text we consider a system in which a subset of modes satisfying energy conservation $\omega_{\kvec_0} = \omega_{\kvec_1} + \omega_{\kvec_2}$ exist. In the high-$Q$ limit $Q_j = \omega_{\kvec_j}/\gamma_j \gg 1$ we find parametric amplification of the energy-conserving mode pair which has the strongest coupling rate $|G|$. In this case the amplified modes oscillate at their natural frequencies $\Omega_j \equiv \omega_{\kvec_j}$. We mention here that besides the spectral tunability of the amplified modes that is demonstrated in the main text, the orientation of the external magnetic field also affects the propagation direction. In Figure~\ref{fig:directionality} we show how the directionality of the amplified magnon modes changes with the orientation of the external magnetic field. \par

To solve for the classical fields above the threshold, we consider again the coupled set of equations. We start by introducing a variable shift to Equations~\eqref{SMeq:class_field_EOM_4} and applying Sokhotski–Plemelj theorem:
\begin{align}
\begin{split}
    0 &= \int_0^\infty d\omega\,  (\Delta_\omega^{(j)}\pm i\epsilon) \left( \alpha_j^*(\omega\pm i\epsilon) \mp \pi \sqrt{\kappa_{j}(\omega\pm i\epsilon)} \beta_j^* \delta(\Delta_\omega^{(j)}) +  i\sqrt{\kappa_{j}(\omega\pm i\epsilon)} \beta_j^*\, \mathcal{P} \biggl( \frac{1}{\Delta_\omega^{(j)}}\biggr) \right)
\end{split}
\end{align}
Here $\mathcal{P}$ represents the Cauchy principal value. Thus, when taking the limit $\epsilon\rightarrow 0$ we are left with a choice of sign which controls whether the coupling of magnons to the detector leads to a decay or amplification term as the solution is plugged back into the other equations of motion. The physical choice is
\begin{align}
    \alpha_j(\omega) = - \pi \left( \sqrt{\kappa_{j}(\omega)} \right)^* \beta_j \delta(\Delta_\omega^{(j)}) -  i\left( \sqrt{\kappa_{j}(\omega)} \right)^* \beta_j\, \mathcal{P} \biggl( \frac{1}{\Delta_\omega^{(j)}}\biggr)
\end{align}
and therefore the term appearing in the remaining equations becomes
\begin{align}
    \int_0^\infty d\omega\, \sqrt{\kappa_{j}(\omega)} \alpha_j(\omega) = -\pi \left| \sqrt{\kappa_{j}(\Omega_j)}\right|^2 \beta_j - i\beta_j \int_0^\infty d\omega\,\mathcal{P} \biggl( \frac{\left| \sqrt{\kappa_{j}(\omega)}\right|^2}{\Delta_\omega^{(j)}}\biggr)
\end{align}
As mentioned below Eq.~\eqref{SMeq:kappa}, treating microwaves as a Markovian bath to the magnons means $\kappa_{j}(\omega)\equiv \kappa_{j}$ over the relevant frequency range. We thus approximate~\cite{gardiner_quantum_2004}
\begin{align}
    \left| \sqrt{\kappa_{j}}\right|^2\int_{-\Omega_j}^\infty d\omega\,\mathcal{P} \biggl( \frac{1}{\omega}\biggr) \approx \left| \sqrt{\kappa_{j}}\right|^2\int_{-\infty}^\infty d\omega\,\mathcal{P} \biggl( \frac{1}{\omega}\biggr) = 0 .
\end{align}
The full steady-state solutions of the classical mean-fields above threshold are therefore
\begin{align}
    |\beta_0|^2 &= \frac{\Delta_1 \Delta_2 + \Gamma_1 \Gamma_2 /4}{|G|^2}\\
    |\beta_1|^2 &= \frac{-i\Delta_2 - \Gamma_2/2}{-|G|^2} \left(\frac{\alpha_\text{d}}{\beta_0} +(-i\Delta_0 - \gamma_0/2)\right)\\
    |\beta_2|^2 &= \frac{-i\Delta_1 - \Gamma_1/2}{-|G|^2} \left(\frac{\alpha_\text{d}}{\beta_0} +(-i\Delta_0 - \gamma_0/2)\right)\\
    \alpha_j(\omega) &= -\pi (\sqrt{\kappa_{j}})^*\beta_j \delta(\Delta_\omega^{(j)}) -i (\sqrt{\kappa_{j}})^*\beta_j \;\mathcal{P}\frac{1}{\Delta_\omega^{(j)}}
\end{align}
where we have introduced $\Gamma_j = \gamma_j + 2\pi |\kappa_{j}|$. We consider the weak-coupling approximation to be valid if $\gamma_j \gg 2\pi |\kappa_{j}|$. Note also that in this weakly coupled system, to reach the long-time limit we require both $\gamma_j\,t\gg1$ and $\kappa_{j}\,t\gg 1$.

\section{Microwave two-mode entanglement}
In this section we outline the calculation of the correlators necessary to quantify microwave two-mode entanglement in the heterodyne detection scheme. We start from the microwave input-output operators 
\begin{align}
    \ain^{(j)} (t) &= \frac{1}{\sqrt{2\pi}} \int d\omega\, e^{-i\Delta_\omega^{(j)}(t-t_0)} \aop_\omega^{(j)}(t_0)\\
    \aout^{(j)} (t) &= \frac{1}{\sqrt{2\pi}} \int d\omega\, e^{-i\Delta_\omega^{(j)}(t-t_0)} \aop_\omega^{(j)}(t_1)
\end{align}
that are related by the equation
\begin{align}\label{SMeq:input_output}
    \aout^{(j)}(t) = \ain^{(j)}(t) -\sqrt{2\pi\kappa_{j}} \bop_j(t).
\end{align}
These operators satisfy the commutation relations 
\begin{gather}
    [\ain^{(i)}(t),\ain^{(j)\dagger}(t^\prime)]= [\aout^{(i)}(t),\aout^{(j)\dagger}(t^\prime)] = \delta_{ij} \delta(t-t^\prime) ,\\
    [\ain^{(i)}(t),\ain^{(j)}(t^\prime)]= [\aout^{(i)}(t),\aout^{(j)}(t^\prime)] = 0 ,\\
    [\ain^{(j)}(t),\aout^{(j^\prime)\dagger}(t^\prime)]= 0 .
\end{gather}
Experimentally, the quantum fluctuations of the two transmission line microwave modes are accessible by a heterodyne detection scheme in which the output field is interfered with a  classical reference field of frequency $\omega_\text{ref}$, so that the detector measures the temporal mode ladder operator 
\begin{align}\label{SMeq:heterodyne_operators}
    \Aop_j\left(\omega_\text{ref}^{(j)}\right) = \int dt\, F_j(t) e^{-i\omega_\text{ref}^{(j)} t} \, \aout^{(j)} (t) \, e^{-i\omega_{\kvec_j}t}  .
\end{align}
The normalized function $F_j(t)$ is the temporal filter function of the detector and the oscillation with frequency $\omega_{\kvec_j}$ arises from the transformation to the original frame. The entanglement between heterodyne detector operators is quantified by the logarithmic negativity 
\begin{align}
    E_N=\max(0,-\ln(2\tilde{\nu}_-))
\end{align}
with $\tilde{\nu}_-$ the smallest symplectic eigenvalue of the partially transposed covariance matrix obtained by flipping the sign of the momentum quadrature of one of the two modes. We thus need the symmetrized covariance matrix $\mathcal{C}_s$ in the basis $\{\hat{x}_1,\hat{p}_1,\hat{x}_2,\hat{p}_2\}$ with
\begin{align}
    \hat{x}_j&=\frac{1}{\sqrt{2}} \left( \Aop_j^\dagger(\omega_\text{ref}^{(j)}) + \Aop_j(\omega_\text{ref}^{(j)}) \right)  \\
    \hat{p}_j&=\frac{i}{\sqrt{2}} \left( \Aop_j^\dagger(\omega_\text{ref}^{(j)}) - \Aop_j(\omega_\text{ref}^{(j)}) \right) 
\end{align}
which can be expressed by $2\times2$-blocks:
\begin{align}
    \mathcal{C}_s = \begin{pmatrix}
        A & C \\ C^T & B
    \end{pmatrix}
\end{align}
Under definition of $\mathcal{D} = |A|+|B|-2|C|$ we can calculate 
\begin{align}
    \tilde{\nu}_- &= \sqrt{\frac{\mathcal{D}-\sqrt{\mathcal{D}^2-4|\mathcal{C}_s|}}{2}} .
\end{align}
\par

Quantifying entanglement requires us to calculate correlators $\langle \hat{\mathbf{V}}_i \hat{\mathbf{V}}_j  \rangle$ of the heterodyne operators Eq.~\eqref{SMeq:heterodyne_operators}, where we define the vector $\hat{\mathbf{V}} = \begin{pmatrix} \Aop_1(\omega_\text{ref}^{(1)}) & \Aop_1^\dagger(\omega_\text{ref}^{(1)}) & \Aop_2(\omega_\text{ref}^{(2)}) & \Aop_2^\dagger(\omega_\text{ref}^{(2)}) \end{pmatrix}^T$. These correlators are calculated analytically. We define the following vectors:
\begin{align}
    \Vopvec_{\IN}(t) = \begin{pmatrix}
        \ain^{(1)}(t) & \ain^{(1)\dagger}(t) & \ain^{(2)}(t) & \ain^{(2)\dagger}(t)
    \end{pmatrix}^T \\
    \Vopvec_{\OUT}(t) = \begin{pmatrix}
        \aout^{(1)}(t) & \aout^{(1)\dagger}(t) & \aout^{(2)}(t) & \aout^{(2)\dagger}(t)
    \end{pmatrix}^T \\
    \vopvec_{\text{m}} = \begin{pmatrix}
        \bop_0(t) & \bop_0^\dagger(t) & \bop_1(t) & \bop_1^\dagger(t) & \bop_2(t) & \bop_2^\dagger(t)
    \end{pmatrix}^T \\
    \mathcal{G} = \begin{pmatrix}
         \sqrt{\kappa_{1}} & (\sqrt{\kappa_{1}})^* & \sqrt{\kappa_{2}} & (\sqrt{\kappa_{2}})^*
     \end{pmatrix}^T
\end{align}
In order to compute the correlations of $\hat{\mathbf{V}}$ we use the input-output relation to split them into three different contributing correlators:
\begin{enumerate}[(i)]
    \item First, the contribution from the input operators describing the thermal noise of the transmission lines. Assuming a thermal state we can approximate the correlations by
    \begin{align}
        \langle \Vopvec_{\IN,i}(t) \Vopvec_{\IN,j}(t+\tau) \rangle = \delta(\tau) \bar{N}_{\text{D},ij}
    \end{align}
    where we have defined the matrix 
    \begin{align}
        \bar{N}_D = \begin{pmatrix}
             0 & (1+\bar{n}_{1}) & 0 & 0 \\
             \bar{n}_{1} & 0 & 0 & 0\\
             0 & 0 & 0 & (1+\bar{n}_{2}) \\
             0 & 0 & \bar{n}_{2} & 0
         \end{pmatrix}
    \end{align} 
    of occupation numbers $\bar{n}_{j}=(\exp(\hbar\omega_{\kvec_j}/(k_B T))-1)^{-1}$ associated with the thermal input state at temperature $T$. This approximation holds for large widths of the temporal filter $F_j$. 
    \item Second, the correlations between transmission line and magnon  fluctuation operators 
    \begin{gather}
        \langle \vopvec_{\text{m},i+2}(t) \Vopvec_{\IN,j}(t+\tau) \rangle\\
        \langle \Vopvec_{\IN,i}(t) \vopvec_{\text{m},j+2}(t+\tau) \rangle 
    \end{gather}
    that arise due to their coupled evolution. To calculate these correlators we formally solve the magnon fluctuation Langevin equations
    \begin{align}
        \frac{d}{dt} \vopvec_{\text{m}}(t) = \bar{M} \vopvec_{\text{m}}(t) + \hat{\mathbf{f}}_L(t) + \hat{\mathbf{c}}(t)
    \end{align}
    where $\hat{\mathbf{f}}_L(t)=(\hat{f}_0(t), \hat{f}_0^\dagger(t), \hat{f}_1(t), \hat{f}_1^\dagger(t), \hat{f}_2(t), \hat{f}_2^\dagger(t))$ is the vector of Langevin force operators related to the magnon decay described by the dissipators in the master equation~\eqref{SMeq:master_eq_collective_preTransform}. The Langevin operator correlations are $\langle \hat{f}_i(t) \hat{f}_j^\dagger(t^\prime) \rangle = \delta_{i,j}\delta(t-t^\prime)\gamma_j(1+\bar{n}_j)$ and the coefficient matrix $\bar{M}$ and operator vector $\hat{\mathbf{c}}(t)$ are 
    \begin{gather}\label{SMeq:appendix_coeff_Matrix}
        \bar{M} = \begin{pmatrix}
            -i\Delta_0 -\frac{\gamma_0}{2} & 0 & -i G\beta_2 & 0 & -iG\beta_1 & 0 \\
            0 & i\Delta_0 - \frac{\gamma_0}{2} & 0 & i G^* \beta_2^* & 0 & i G^* \beta_1^* \\
            -iG^* \beta_2^* & 0 & -i\Delta_1-\frac{\Gamma_1}{2} & 0 & 0 & -iG^* \beta_0 \\
            0 & iG\beta_2 & 0 & i\Delta_1-\frac{\Gamma_1}{2} & iG\beta_0^* & 0 \\
            -iG^* \beta_1^* & 0 & 0 & -iG^*\beta_0 & -i\Delta_2-\frac{\Gamma_2}{2} & 0 \\
            0 & iG\beta_1 & iG\beta_0^* & 0 & 0 & i\Delta_2-\frac{\Gamma_2}{2}
            \end{pmatrix} ,\\
        \hat{\mathbf{c}}(t) = \begin{pmatrix}
            0 \\ 
            0 \\
            \sqrt{2\pi} \sqrt{\kappa_{1}} \ain^{(1)} (t)\\
            \sqrt{2\pi} (\sqrt{\kappa_{1}})^* \ain^{(1)\dagger}(t)\\
            \sqrt{2\pi} \sqrt{\kappa_{2}} \ain^{(2)}(t)\\
            \sqrt{2\pi} (\sqrt{\kappa_{2}})^* \ain^{(2)\dagger}(t)
        \end{pmatrix} .
    \end{gather}
    By integration from a past time $\tau_0$ one finds the formal solution
    \begin{align}\label{SMeq:formal_sol}
        \vopvec_{\text{m}} (t) = e^{\bar{M}(t-\tau_0)} \vopvec_{\text{m}} (\tau_0) + \int_{\tau_0}^t d\tau\, e^{\bar{M}(t-\tau)} (\hat{\mathbf{f}}_L(\tau) + \hat{\mathbf{c}}(\tau) )
    \end{align}
    We assume that at initial time $\tau_0$ the microwave input operators and magnon operators are uncorrelated $\langle \Vopvec_{\IN,i}(t) \vopvec_{\text{m},j+2}(\tau_0) \rangle = 0$.  Furthermore, magnon Langevin noise operators are uncorrelated with input operators at all times, i.e., $\langle \Vopvec_{\IN,i}(t) \hat{\mathbf{f}}_{L,j}(t^\prime) \rangle =0$ and, since we are considering the long-time limit, the correlators are invariant under time translation. As a result we find 
    \begin{align}
        \langle \vopvec_{\text{m},i+2}(t) \Vopvec_{\IN,j}(t+\tau) \rangle = \sqrt{2\pi} \mathcal{G}_i \sum_{\alpha=1}^4 (e^{\bar{M}(-\tau)})_{(i+2)(\alpha+2)} \bar{N}_{\text{D},\alpha j}
    \end{align}
    for $\tau \leq 0$ and zero otherwise, and 
    \begin{align}
        \langle \Vopvec_{\IN,i}(t) \vopvec_{\text{m},j+2}(t+\tau) \rangle = \sum_{\alpha=1}^4 \sqrt{2\pi} \mathcal{G}_j (e^{\bar{M}\tau})_{(j+2)(\alpha+2)} \bar{N}_{\text{D},i\alpha}
    \end{align}
    for $\tau \geq 0$ and zero otherwise.
    \item Finally we calculate the magnon correlators
    \begin{align}
        \langle \vopvec_{\text{m},i}(t) \vopvec_{\text{m},j}(t+\tau) \rangle
    \end{align}
    for $i,j\in\{1,\dots,6\}$. First we need to derive the steady-state equal time correlators $\Sigma_{i,j} = \langle \vopvec_{\text{m},i}(t) \vopvec_{\text{m},j}(t) \rangle_{ss}$. The dynamics are governed by
    \begin{align}
            0= \frac{d}{dt} \langle \vopvec_{\text{m},i}(t) \vopvec_{\text{m},j}(t) \rangle = \bar{M}_{i \alpha} \Sigma_{\alpha j} + \Sigma_{i\alpha} \bar{M}_{\alpha j}+ \langle \hat{\mathbf{f}}_{L,i} \vopvec_{\text{m},j} \rangle + \langle \vopvec_{\text{m},i} \hat{\mathbf{f}}_{L,j} \rangle + \langle \hat{\mathbf{c}}_{i} \vopvec_{\text{m},j} \rangle + \langle \vopvec_{\text{m},i} \hat{\mathbf{c}}_{j} \rangle 
    \end{align}
    with Einstein summation convention for index $\alpha$. We use the above derived formal solution of $\vopvec_{\text{m}}$ (Eq.~\eqref{SMeq:formal_sol}) to obtain the following correlations:
    \begin{gather}
        \langle \vopvec_{\text{m},i}(t) \hat{\mathbf{f}}_{L,j}(t) \rangle = \frac{1}{2}\begin{pmatrix}
            0 & \gamma_0 (1+\bar{n}_0) & 0 & 0 & 0 & 0 \\
            \gamma_0 \bar{n}_0 & 0 & 0 & 0 & 0 & 0 \\
            0 &  0 & 0 & \gamma_1 (1+\bar{n}_1) & 0 & 0 \\
            0 & 0 & \gamma_1 \bar{n}_1 &0 & 0 & 0 \\
            0 & 0 & 0 & 0 & 0 & \gamma_2 (1+\bar{n}_2) \\
            0 & 0 & 0 & 0 & \gamma_2 \bar{n}_2 & 0 \\
        \end{pmatrix}\\
        \langle \hat{\mathbf{c}}_{i} \vopvec_{\text{m},j} \rangle = 2\pi \mathcal{G}_{i-2} \mathcal{G}_{j-2} \begin{pmatrix}
             0&0& 0 & 0 & 0 & 0 \\
             0&0& 0 & 0 & 0 & 0 \\
             0&0& 0 & (1+\bar{n}_{1}) & 0 & 0 \\
             0&0& \bar{n}_{1} & 0 & 0 & 0\\
             0&0& 0 & 0 & 0 & (1+\bar{n}_{2}) \\
             0&0& 0 & 0 & \bar{n}_{2} & 0
         \end{pmatrix}_{i,j}
    \end{gather}
    Using the diagonalizability of $\bar{M}=PDP^{-1}$ the equation can be solved as
    \begin{align}
        \Sigma_{ij} = \sum_{\alpha,\beta,\gamma,\delta=1}^6 P_{i\alpha} \frac{-P_{\alpha \gamma}^{-1} C_{\gamma\delta} P_{\beta \delta}^{-1}}{D_{\alpha\alpha}+D_{\beta\beta}} P_{j\beta}
    \end{align}
    with $C_{i,j}=\langle \hat{\mathbf{f}}_{L,i} \vopvec_{\text{m},j} \rangle + \langle \vopvec_{\text{m},i} \hat{\mathbf{f}}_{L,j} \rangle + \langle \hat{\mathbf{c}}_{i} \vopvec_{\text{m},j} \rangle + \langle \vopvec_{\text{m},i} \hat{\mathbf{c}}_{j} \rangle $, allowing us to calculate the unequal-time correlators using the quantum regression theorem. From $\frac{d}{dt} \langle \vopvec_{\text{m}} (t) \rangle = \bar{M} \langle \vopvec_{\text{m}} (t) \rangle$ we get
    \begin{align}
        \langle \vopvec_{\text{m},i}(0) \vopvec_{\text{m},j}(\tau) \rangle &= (e^{\bar{M}\tau})_{j\alpha } \Sigma_{i \alpha} \\
        \langle \vopvec_{\text{m},i}(\tau) \vopvec_{\text{m},j}(0) \rangle &= (e^{\bar{M}\tau})_{i\alpha } \Sigma_{\alpha j}
    \end{align}
    for positive times $\tau \geq 0$. 
\end{enumerate}
The full analytical expression for the covariance matrix, while straightforward to assemble from the above correlators, is cumbersome and we evaluate the logarithmic negativity numerically.

\end{document}